\documentclass[twocolumn,prd,floatfix,preprintnumbers,nofootinbib,superscriptaddress,]{revtex4-2}

\usepackage{bm}
\usepackage{times}
\usepackage{amssymb,amsbsy,amsmath,amsfonts}
\usepackage{graphicx}
\usepackage{float}
\usepackage{color}
\usepackage{morefloats}
\usepackage{rotating}
\usepackage{srcltx}
\usepackage{slashed}
\usepackage{subfigure}
\usepackage{multirow,diagbox}
\usepackage{verbatim}
\usepackage{hyperref}
\usepackage{tabularx}
\usepackage[outdir=./]{epstopdf}
\usepackage[normalem]{ulem}
\hypersetup{pdftex,colorlinks=true,linkcolor=blue,citecolor=blue,menucolor=black,urlcolor=black,filecolor=blue}
\begin{document}

\title{Theoretical study on $\Lambda_c^+ \to \Lambda K^+\bar{K}^0$ decay and $\Xi^*(1690)$ resonance}

\author{Si-Wei Liu}~\email{liusiwei@impcas.ac.cn}
 \affiliation{Institute of Modern Physics, Chinese Academy of
Sciences, Lanzhou 730000, China} 
\affiliation{School of Nuclear Sciences and Technology, University of Chinese Academy of Sciences, Beijing 101408, China}
\author{Qing-Hua Shen}~\email{shenqinghua@impcas.ac.cn}
 \affiliation{Institute of Modern Physics, Chinese Academy of Sciences, Lanzhou 730000, China} 
\affiliation{School of Nuclear Sciences and Technology, University of Chinese Academy of Sciences, Beijing 101408, China}
\author{Ju-Jun Xie}~\email{xiejujun@impcas.ac.cn}
 \affiliation{Institute of Modern Physics, Chinese Academy of
Sciences, Lanzhou 730000, China} 
\affiliation{School of Nuclear Sciences and Technology, University of Chinese Academy of Sciences,
Beijing 101408, China}
\affiliation{Southern Center for Nuclear-Science Theory (SCNT), Institute of Modern Physics, Chinese Academy of Sciences, Huizhou 516000, Guangdong Province, China}

\date{\today}

\begin{abstract}

We present a theoretical study of $\Xi^*(1690)$ resonance in the $\Lambda_c^+ \to \Lambda K^+ \bar{K}^0$ decay, where the weak interaction part proceeds through the Cabibbo-favored process $c \to s + u\bar{d}$. Next, the intermediate two mesons and one baryon state can be constructed with a pair of $q\bar{q}$ with the vacuum quantum numbers. Finally, the $\Xi^*(1690)$ is mainly produced from the final state interactions of $\bar{K}\Lambda$ in coupled channels, and it is shown in the $\bar{K}\Lambda$ invariant mass distribution. Besides, the scalar meson $a_0(980)$ and nucleon excited state $N^*(1535)$ are also taken into account in the decaying channels $K^+\bar{K}^0$ and $K^+\Lambda$, respectively. Within model parameters, the $K^+ \bar{K}^0$, $\bar{K}^0 \Lambda$ and $K^+ \Lambda$ invariant mass distributions are calculated, and it is found that our theoretical results can reproduce well the experimental measurements, especially for the clear peak around $1690$ MeV in the $\bar{K}\Lambda$ spectrum. The proposed weak decay process $\Lambda_c^+ \to \Lambda K^+ \bar{K}^0$ and the interaction mechanism can provide valuable information on the nature of the $\Xi^*(1690)$ resonance.

\end{abstract}

\maketitle

\section{Introduction} \label{sec:Introduction}

The study of the properties of the $\Xi^*$ states concentrates much attention in hadron physics~\cite{Garcia-Recio:2003ejq,Gamermann:2011mq,Pervin:2007wa,PavonValderrama:2011gp,Nishibuchi:2022zfo}. However, our knowledge about the $\Xi$ spectrum is quite scarce~\cite{ParticleDataGroup:2020ssz}. Except for the ground $\Xi(1321)$ state with spin-parity $J^P = 1/2^+$ and $\Xi^*(1530)$ with $J^P = 3/2^+$ being well established with four-star ratings, the situation of other $\Xi^*$ excited states is still rather uncertain with less than three-star ratings~\cite{ParticleDataGroup:2020ssz}. For some of them, their existence has not been confirmed. Hence, further studies of the $\Xi^*$ resonances both on the experimental and theoretical sides are necessary~\cite{Nagae:2005rj,Guo:2007dw,Schumacher:2010zz,PANDA:2009yku}.

For the study of $\Xi^*$ resonances, the $\bar{K}N$ scattering is available~\cite{Ahn:2018hbc}. Indeed, the $\Xi^*(1690)$ resonance was first observed in the $\bar{K}\Sigma$ invariant mass spectrum~\cite{Amsterdam-CERN-Nijmegen-Oxford:1978chc} in the $K^-p$ reactions at $4.2$ GeV. In the measurements of Ref.~\cite{Amsterdam-CERN-Nijmegen-Oxford:1978chc}, the mass and total decay width of $\Xi^*(1690)$ are established as $M=1694\pm6$ MeV and $\Gamma=26\pm6$ MeV in the charged channel and $M=1684\pm5$ MeV and $\Gamma=20 \pm 4$ MeV in the neutral channel. In Ref.~\cite{WA89:1997vxx}, the $\Xi^*(1690)$ resonance was confirmed by the WA89 collaboration in the neutral $\pi^+\Xi^-$ channel with measured mass $M = 1686 \pm 4$ MeV and width $\Gamma = 10 \pm 
6$ MeV. In Ref.~\cite{BaBar:2008myc}, in addition to the investigations of the $\Xi^*(1530)$ properties in the $\Lambda^+_c \to K^+\pi^+\Xi^-$ decay process, evidence of the existence of the $\Xi^*(1690)$ resonance with $J^P= 1/2^-$ was also found by $\it BABAR$ collaboration. In Ref.~\cite{Belle:2001hyr}, the Belle collaboration presented the first evidence for the process $\Lambda^+_c \to K^+ \Xi^{*0}(1690) \to K^+K^-\Sigma^+$, and the fitted mass and width of $\Xi^*(1690)$ are $M=1688 \pm 2$ MeV and $\Gamma= 11 \pm 4$ MeV. Furthermore, contribution of the $\Xi^{*0}(1690)$ to the $\Lambda^+_c \to K^+ \Xi^{*0}(1690) \to K^+ \bar{K}^0\Lambda$ reaction was also found~\cite{Belle:2001hyr}. In Ref.~\cite{BaBar:2006tck}, the $\Xi(1690)$ resonance was also observed in the $\bar{K}^0\Lambda$ channel in the decay $\Lambda^+_c \to K^+ \bar{K}^0\Lambda$ by the ${\it BABAR}$ collaboration, where a coherent amplitude analysis of the $\Lambda a_0(980)$ and $K^+\Xi^{*0}(1690)$ productions was performed, the obtained mass and width of $\Xi^*(1690)$ resonance are $M=1684.7\pm1.3(\text{stat.})_{-1.6}^{+2.2}(\text{syst.})$ MeV, $\Gamma=8.1_{-3.5}^{+3.9}(\text{stat.})_{-0.9}^{+1.0}(\text{syst.})$ MeV, and its spin is consistent with $1/2$. Furthermore, the $\Xi^*(1690)$ resonance has also been found in the hyperon-nucleon interactions~\cite{Biagi:1981cu,Biagi:1986zj}. In most of these experimental analysis the spin-parity of $\Xi^*(1690)$ resonance favor $J^P = 1/2^-$.

In 2019, the $\Xi^*(1620)$ resonance was observed in the $\pi^+\Xi^-$ invariant spectrum via the $\Xi_c^+ \to \Xi^-\pi^+\pi^+$ decay. Meanwhile, the evidence of the $\Xi^*(1690)$ resonance was also found with the same data sample, and its significance is $4\sigma$~\cite{Belle:2018lws}. However, up to present time the quantum numbers of $\Xi^*(1690)$ have not yet been determined. Nevertheless, to fully understand the nature of $\Xi^*(1690)$ resonance, further experiments are certainly required.

On the theoretical side, within the constituent quark models, the first excited state of $\Xi$ baryon is around $1800$ MeV~\cite{Chao:1980em,Capstick:1986ter,Xiao:2013xi}, thus it is difficult to treat $\Xi^*(1690)$ as a three-quark state. In Ref.~\cite{Pervin:2007wa}, the $\Xi^*(1690)$ was treated as a three-quark state and its spin-parity quantum numbers are $J^P = 1/2^-$. Note that the obtained mass with the constituent quark model~\cite{Pervin:2007wa}, is 1725 MeV, which is still 35 MeV higher than its nominal mass. This implies that $\Xi^*(1690)$ might have some nontrivial structure other than the usual three-quark state. In fact, using the chiral unitary approach, the $\Xi^*(1690)$ can be interpreted as an $s$-wave meson-baryon molecular state~\cite{Garcia-Recio:2003ejq,Gamermann:2011mq,Sekihara:2015qqa}. It couples strongly to the $\bar{K}\Sigma$ channel~\cite{Sekihara:2015qqa}, and its coupling to the $\pi\Xi$ channel is small. Thus its narrow width can be naturally explained~\cite{Sekihara:2015qqa}. Furthermore, its spin-parity are $1/2^-$ in the chiral unitary approach. In Ref.~\cite{Aliev:2018hre}, the $\Xi^*(1690)$ was investigated by means of the two-point QCD sum rules, where it was also concluded that the $\Xi^*(1690)$ state, most probably, has quantum numbers spin-parity $J^P = 1/2^-$. However, the obtained width for the $\Xi^*(1690)$ state is about $100$ MeV~\cite{Aliev:2018hre}, which is much larger than the experimental measurements~\cite{ParticleDataGroup:2020ssz}.

The main property of $\Xi^*(1690)$ resonance is that its decay width is too narrow compared with other baryon resonances that have similar mass~\cite{ParticleDataGroup:2020ssz}. In spite of $\Xi^*(1690)$ state has a large phase space to decay to open channels, such as $\pi\Xi$ and $\bar{K}\Lambda$, its width is just in the order of $10$ MeV. In Ref.~\cite{Khemchandani:2016ftn}, using the chiral unitary approach, the $\Xi^*(1690)$ state is dynamically generated in the pseudoscalar-baryon and vector-baryon coupled channels~\footnote{Note that the transitions between pseudoscalar-baryon and vector-baryon channels are crucial to obtain the pole for $\Xi^*(1690)$ state. If those transitions were zero, the pole of $\Xi^*(1690)$ disappears.}. It was found that most of the properties of $\Xi^*(1690)$, especially its narrow width, can be well explained~\cite{Khemchandani:2016ftn}. In that work, the $\Xi^*(1690)$ has strong couplings to $\bar{K}\Sigma$ and $\eta \Xi$ channels, while its couplings to the $\bar{K}\Lambda$ and $\pi\Xi$ channels are small. Recently, the meson-baryon interaction in the neutral strangeness $S=-2$ sector was studied using an extended Unitarized Chiral Perturbation Theory, where the leading Weinberg-Tomozawa term, Born term, and the next-to-leading order contributions are considered~\cite{Feijoo:2023wua}. It was found that both $\Xi^*(1620)$ and $\Xi^*(1690)$ states can be dynamically generated, and the obtained properties of them are in reasonable agreement with the known experimental data.

It has been shown that the weak decay of charmed baryons governed by $c \to s$ quark transition is beneficial to probe the strange baryons, some of them are subjects of intense debate about their nature~\cite{Oset:2016lyh,Oset:2016nvf,Yu:2020vlt}. For instance, the hyperon production from the $\Lambda^+_c \to K^- p \pi^+$ and $\Lambda^+_c \to K^0_S p \pi^0$ decays were investigated within the effective Lagrangian approach in Ref.~\cite{Ahn:2019rdr}. The $\Xi^*(1620)$ and $\Xi^*(1690)$ states were studied in the $\Xi_c \to \pi^+ MB$ process in Ref.~\cite{Miyahara:2016yyh}, where $M$ and $B$ stand for mesons and baryons, respectively. It was shown that these weak decays might be an ideal tool to study the $\Xi^*(1620)$ and $\Xi^*(1690)$ resonances, which are dynamically produced in the rescattering of $M$ and $B$ in the final states.  

Therefore, in this work, we take advantage of those ideas and revisit the $\Lambda_c^+ \to \Lambda K^+ \bar{K}^0$ decay, which requires the creation of an $s\bar{s}$ quark pair. Following Refs.~\cite{Sekihara:2015qqa,Khemchandani:2016ftn,Miyahara:2016yyh}, in addition to the contributions of the $\Xi^*(1690)$ resonance that is dynamically generated from the final state interactions of $\bar{K}^0\Lambda$ in $S$ wave, we will consider also the contributions of the $a_0(980)$ and $N^*(1535)$ states, where they are dynamically produced by the final state interactions of $K^+\bar{K}^0$~\cite{Nieves:1998hp,Oller:1997ti} and $\Lambda K^+$~\cite{Kaiser:1995cy,Nieves:2001wt,Inoue:2001ip,Bruns:2010sv,Khemchandani:2013nma,Garzon:2014ida} in coupled channels, respectively.

This article is organized as follows. In the next section, we present the theoretical formalism for studying the $\Lambda_c^+\to\Lambda K^+\bar{K}^0$ decay. In Sec.~\ref{sec:Results}, theoretical numerical results and discussions and presented, followed by a short summary in Sec.~\ref{sec:Summary}.

\section{Formalism} \label{sec:Formalism}

For the production of $\Xi^*(1690)$ in the weak decay process $\Lambda^+_c \to K^+ \bar{K}^0 \Lambda$, we firstly take it as a dynamically generated state in the final state interaction of $\bar{K}\Lambda$ in coupled channels. Secondly, we take $\Xi^*(1690)$ as a Breit-Wigner resonance. The roles of $a_0(980)$ and $N^*(1535)$ are also investigated, where the contribution of $a_0(980)$ state in encoded in the $s$-wave $K\bar{K}$ final state interaction as done in Refs.~\cite{Xie:2014tma,Ling:2021qzl,Wang:2021naf,Molina:2019udw,Duan:2020vye,Zhu:2022duu}, and the one of the $N^*(1535)$ resonance is in the $s$-wave $K\Lambda$ final state interaction~\cite{Kaiser:1995cy,Nieves:2001wt,Inoue:2001ip,Bruns:2010sv,Khemchandani:2013nma,Garzon:2014ida}.

\begin{figure}[htbp]
\includegraphics[scale=0.8]{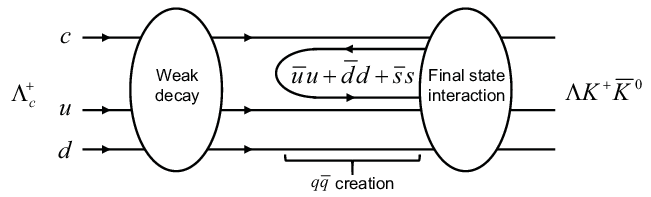}
\caption{The schematic diagram for the Cabibbo-favored process $\Lambda_c^+\to\Lambda K^+\bar{K}^0$.}
\label{fig:w-d-sche-diag}
\end{figure}

The schematic diagram for the Cabibbo-favored process $\Lambda_c^+ \to \Lambda K^+\bar{K}^0$ is presented in Fig.~\ref{fig:w-d-sche-diag}, where the decay process proceeds in three parts:
\begin{itemize}
    \item In the first step, the $c$ quark in $\Lambda^+_c$ turns into an $s$ quark via the weak decay. 
    \item Introduce a $q\bar{q}$ pair with the quantum numbers of the vacuum to form a pseudoscalar-baryon ($PB$) or pseudoscalar-pseudoscalar ($PP$) pair. 
    \item The final state interactions are taken into account in coupled channels within the chiral unitary approach, which will lead to the dynamical productions of the $a_0(980)$, $N^*(1535)$, and $\Xi^*(1690)$ resonances.
\end{itemize}

\subsection{Contributions of the $a_0(980)$, $N^*(1535)$, and $\Xi^*(1690)$ resonances from final state interactions}

\begin{figure*}[htbp]
\includegraphics[scale=1.2]{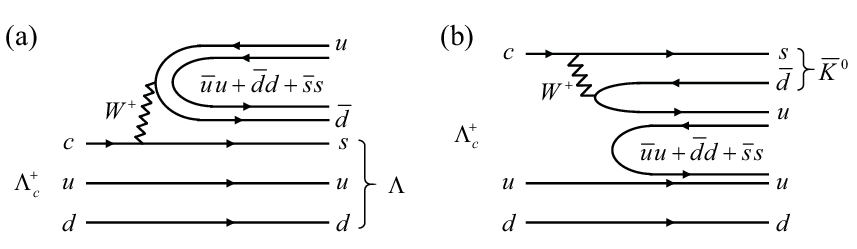}
\caption{The decay mechanisms for (a): $\Lambda_c^+ \to \Lambda + (PP)^+$, (b): $\Lambda_c^+ \to \bar{K}^0 + (PB)^+$}
\label{fig:a-w-d-p}
\end{figure*}

\begin{figure*}[htbp]
    \centering
    \includegraphics[scale=1.2]{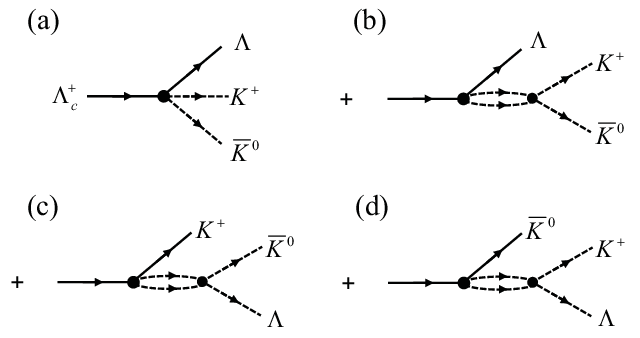}
    \caption{The schematic diagram for the final state interaction for the $\Lambda_c^+\to\Lambda K^+\bar{K}^0$ decay. (a): tree level diagram, (b): $K^+\bar{K}^0$ rescattering, (c): $\bar{K}^0\Lambda$ rescattering, and (d): $K^+\Lambda$ rescattering.}
    \label{fig:final-s-i}
\end{figure*}

Following Refs.~\cite{Miyahara:2016yyh,Xie:2016evi,Xie:2017erh,Xie:2017xwx,Wang:2022nac,Pavao:2018wdf}, we assume two primary decay mechanisms for the process $\Lambda_c^+ \to \Lambda K^+ \bar{K}^0$, which are presented in Fig.~\ref{fig:a-w-d-p}. In the weak decay of charmed hadrons the diagrams are classified in six different topologies~\cite{Chau:1987tk,Chau:1990ij,Chau:1995gk,Qin:2022nof}: $W$ external emission, $W$ internal emission, $W$-exchange, $W$-annihilation, horizontal $W$-loop and vertical $W$-loop. Here, we consider the dominant $W$ external emission and internal emission diagrams as shown in Fig.~\ref{fig:a-w-d-p} (a) and (b), respectively. The $c$ quark in $\Lambda^+_c$ turns into an $s$ quark and a $W^+$ boson, followed by the $W^+$ boson decaying into a $u \bar{d}$ pair. To get the final states $\Lambda K^+ \bar{K}^0$, the $sud$ ($s \bar{d}$) cluster forms the $\Lambda$ ($\bar{K}^0$), and the $u \bar{d}$ ($uud$), together with the $q \bar{q}$ ($=u\bar{u} + d\bar{d} + s\bar{s}$) pair created from the vacuum, hadronize into $K^+ \bar{K}^0$ ($K^+\Lambda$). In the former processes, the $ud$ diquark in the $\Lambda^+_c$ is always kept. Therefore, the $\Lambda_c^+$ weak decay processes shown in Fig.~\ref{fig:a-w-d-p} can be formulated as following~\cite{Xie:2016evi,Xie:2017erh,Xie:2017xwx,Wang:2022nac,Pavao:2018wdf}:
\begin{align}
\Lambda_{c}^{+}&=\frac{1}{\sqrt{2}}\left|c(ud-du)\right\rangle,\nonumber\\
&\to-\frac{\sqrt{6}}{3}V_1\left|\Lambda\right\rangle\left| u(u\bar{u}+d\bar{d}+s\bar{s})\bar{d}\right\rangle,
\label{eq:K-K-w-d}\\
&\to\frac{1}{\sqrt{2}}V_2\left|\bar{K}^0\right\rangle\left| u(u\bar{u}+d\bar{d}+s\bar{s})(ud-du)\right\rangle,
\label{eq:K+-lam-w-d}
\end{align}
where $V_1$ and $V_2$ are the strength of the production vertices~\footnote{We have assumed that these production vertices proceed through $S$-wave interaction.}, which contain all the dynamical factors. According to these results of Ref.~\cite{Miyahara:2016yyh}, we then connect two degrees of freedom, the quarks and the hadons. Then we can rewrite the intermediate states as
\begin{align}
    \Lambda_{c}^{+}&\to V_1\left|\Lambda\right\rangle\left|-\frac{2\sqrt{2}}{3}\pi^+\eta -\frac{\sqrt{6}}{3} K^+\bar{K}^0\right\rangle \label{eq:K-K-P-P}\\
    &\to V_2\left|\bar{K}^0\right\rangle\left|(\frac{\eta}{\sqrt{3}}+\frac{\pi^0}{\sqrt{2}})p+\pi^+n-\frac{\sqrt{6}}{3}K^+\Lambda\right\rangle, \label{eq:K+-lam-P-B}
\end{align}
where we have omitted the $\eta'$ terms as in Refs.~\cite{Miyahara:2016yyh,Sekihara:2015qqa,Khemchandani:2016ftn}, because its mass threshold is located much higher in the energy and its contribution should be small. After the production of the $PP$ or $PB$ pair, the final state interactions between intermediate mesons and baryons come up, which is shown in Fig.~\ref{fig:final-s-i}. And, we can obtain the explicit expressions of the decay amplitude $\mathcal{M}_i$ ($i =a$, $b$, $c$, and $d$ according to these diagrams shown in Fig.~\ref{fig:final-s-i}) using Eqs.~\eqref{eq:K-K-P-P} and \eqref{eq:K+-lam-P-B}:
\begin{eqnarray}
        \mathcal{M}_a &=& \mathcal{M}^{\rm Tree} = -\frac{\sqrt{6}}{3}(V_1 + V_2),\\        \mathcal{M}_b &=& \mathcal{M}^{\rm FSI}_{K^+ \bar{K}^0} = h_{K^+\bar{K}^0}G_{K^+\bar{K}^0}T_{K\bar{K}\to K\bar{K}}^{I=1} \nonumber \\
        && -h_{\pi^+\eta}G_{\pi^+\eta}T_{\pi\eta\to K\bar{K}}^{I=1}, \\ 
        \mathcal{M}_c &=& \mathcal{M}^{\rm FSI}_{\bar{K}^0\Lambda} = h_{\bar{K}^0\Lambda}G_{\bar{K}^0\Lambda}T_{\bar{K}\Lambda\to \bar{K}\Lambda}^{I=1/2}, \\
        \mathcal{M}_d &=& \mathcal{M}^{\rm FSI}_{K^+\Lambda} = h_{K^+\Lambda}G_{K^+\Lambda}T_{K\Lambda\to K\Lambda}^{I=1/2} \nonumber \\
        && -(\sqrt{\frac{2}{3}}h_{\pi^+n}G_{\pi^+n}+\frac{1}{\sqrt{3}}h_{\pi^0p}G_{\pi^0p})T_{\pi N\to K\Lambda}^{I=1/2} \nonumber \\
        && +h_{\eta p}G_{\eta p}T_{\eta N\to K\Lambda}^{I=1/2}, \label{eq:MK+lambda}
\end{eqnarray}
where $G_{PP}$ and $G_{PB}$ are the loop functions of the $PP$ or the $PB$ propagator, respectively. The coefficients $h_{PP}$ and $h_{PB}$ for the terms of final state interactions are
\begin{eqnarray}
        h_{K^+\bar{K}^0} &=& -\frac{\sqrt{6}}{3}(V_1+V_2), ~
        h_{\pi^+\eta}  = -\frac{2\sqrt{2}}{3}V_1,\\
        h_{\bar{K}^0\Lambda} &=& h_{K^+\Lambda} = -\frac{\sqrt{6}}{3}(V_1+V_2), \\
        h_{\pi^+n} &=& V_2,~
        h_{\pi^0p} =\frac{\sqrt{2}}{2}V_2, ~
        h_{\eta p} =\frac{\sqrt{3}}{3}V_2. \label{eq:hcoeff}
\end{eqnarray}

The analytic form of the loop function $G$, with dimensional regularization, is given by
\begin{align}
	G(s) = \frac{2M}{16\pi^2}&\Bigg\{a_{\mu} + \text{ln}\frac{M^2}{\mu^2}+\frac{m^2-M^2+s}{2s}\text{ln}\frac{m^2}{M^2} \nonumber \\
	&+\frac{p}{\sqrt{s}}\bigg[\text{ln}\left(s-\left(m^2-M^2\right)+2p\sqrt{s}\right)\nonumber \\
	&+\text{ln}\left(s+\left(m^2-M^2\right)+2p\sqrt{s}\right)\nonumber \\
	&-\text{ln}\left(-s+\left(m^2-M^2\right)+2p\sqrt{s}\right)\nonumber \\
	&-\text{ln}\left(-s-\left(m^2-M^2\right)+2p\sqrt{s}\right)\bigg]\Bigg\},
\end{align}
where $\mu$ is a scale of dimensional regularization, and $a_{\mu}$ is the subtraction constant. Any change of $\mu$ can be reabsorbed by a change in $a_\mu$. In this work, we choose $\mu = 630$ MeV, while $a_\mu$ will be determined from the experimental data, and they will be discussed in the following. $m$ and $M$ are the masses of the meson and baryon in the loop, respectively. $p$ represents the magnitude of the three-momentum of one particle in the meson-meson or meson-baryon rest frame,

\begin{align}
	p&=\frac{\lambda^{1/2}(s,m^2,M^2)}{2\sqrt{s}}, \\
	\lambda^{1/2}(x,y,z)&=\sqrt{x^2+y^2+z^2-2xy-2yz-2xz}.
\end{align}

The two body scattering amplitudes $T_{PP \to PP}$ and $T_{PB \to PB}$ are obtained by solving the Bethe-Salpeter equation within the chiral unitary approach~\cite{Oller:1997ti,Oller:1997ng,Oller:1998hw}
\begin{equation}
    T=[1-VG]^{-1}V,
    \label{eq:BS-equ}
\end{equation}
where $V$ is the transition potential between the involved channels, which are given explicitly in Refs.~\cite{Oller:1997ti,Garzon:2014ida,Khemchandani:2016ftn,Garzon:2016evl}. The scattering amplitude element $T$ in Eq.~\eqref{eq:BS-equ} in the particle basis can be related to the one in the isospin basis, and we get
\begin{eqnarray}
T_{K^+\bar{K}^0\to K^+\bar{K}^0} &=& T_{K\bar{K}\to K\bar{K}}^{I=1}, \\
T_{\pi^+\eta\to K^+\bar{K}^0} &=& -T_{\pi\eta\to K\bar{K}}^{I=1}, \\
T_{\bar{K}^0\Lambda\to \bar{K}^0\Lambda} &=& T_{\bar{K}\Lambda\to \bar{K}\Lambda}^{I=1/2}, \\
T_{K^+\Lambda\to K^+\Lambda} &=& T_{K\Lambda\to K\Lambda}^{I=1/2}, \\
T_{\pi^+n\to K^+\Lambda} &=& -\sqrt{\frac{2}{3}}T_{\pi N\to K\Lambda}^{I=1/2},  \\
T_{\pi^0p\to K^+\Lambda} &=& -\sqrt{\frac{1}{3}}T_{\pi N\to   K\Lambda}^{I=1/2}, \\ 
T_{\eta p\to K^+\Lambda} &=& T_{\eta N\to K\Lambda}^{I=1/2}.
        \label{equ:charge-isospin}
\end{eqnarray}
 
Although the form of these $K\bar{K}$, $K\Lambda$, and $\bar{K}\Lambda$ interactions have been detailed elsewhere~\cite{Oller:1997ti,Inoue:2001ip,Ramos:2002xh,Garzon:2014ida,Khemchandani:2016ftn,Garzon:2016evl}, we
briefly revisit here the $\bar{K} \Lambda$ case. This will allow us to review the general procedure of calculating the two-body amplitudes entering the total decay amplitude of $\Lambda^+_c \to \Lambda \bar{K}^0 K^+$ reaction.

In order to describe the $\Xi^*(1690)$ state, a coupled channel analysis was performed. In the isospin $1/2$ sector, there are four coupled channels, which are: $\pi \Xi$, $\bar{K}\Lambda$, $\bar{K}\Sigma$, and $\eta \Xi$. These channels are labeled by the indices $j=1$, ..., $4$. For the transition potential $V_{ij}$, it is expressed as~\cite{Ramos:2002xh,Khemchandani:2016ftn}
\begin{eqnarray}
V_{ij} & = & -\frac{C_{ij}}{4f_i f_j}(2\sqrt{s}-M_i-M_j) \nonumber \\ 
&& \times \sqrt{\frac{(M_i+E_i)(M_j+E_j)}{4M_iM_j}},
\end{eqnarray}
where $f_i$ is the meson decay constant of the $i$th channel, $f_\pi = 92.1$ MeV, $f_K=1.2f_\pi$ and $f_\eta=1.3f_\pi$. $E_i$ and $E_j$ are the initial and final baryon energies, and $E_i = \sqrt{M_i^2+|\vec{p}_i|^2}$, $|\vec{p}_i|=\frac{\lambda^{1/2}(s,m_i^2,M_i^2)}{2\sqrt{s}}\theta(\sqrt{s}-M_i-m_i)$, with $s$ the invariant mass squared of the meson-baryon system, and the meson and baryon masses $m_i$ and $M_i$ in the $i$th channel, respectively. The factor $C_{ij}$ is symmetric and its value is listed in Table~\ref{tab:cij}. 

\begin{table}[htbp]
	\begin{center}
\renewcommand\arraystretch{1.5}	
\caption{Coefficients $C_{ij}$ for $\pi \Xi$, $\bar{K}\Lambda$, $\bar{K}\Sigma$, and $\eta \Xi$ coupled channels in the isospin $I=1/2$ basis.}
		\begin{tabular}{ccccc}
		\hline \hline
			 & $\pi \Xi$  & $\bar{K} \Lambda$  & $\bar{K} \Sigma$ & $\eta \Xi$    \\ \hline
	$\pi \Xi$  & $2$     & $-3/2$   & $-1/2$   & $0$ \\ 
$\bar{K} \Lambda$   & & $0$ & $0$ & $-3/2$   \\ 
        $\bar{K}\Sigma$ & & & $2$ & $3/2$ \\
		$\eta \Xi$  &  &   &    & $0$  \\   	\hline \hline
		\end{tabular} \label{tab:cij}
	\end{center}
\end{table}

Then one can solve the Bethe-Salpeter equation as shown in Eq.~\ref{eq:BS-equ} with the on-shell factorized potential and, thus, the $S$ wave scattering $T_{ij}$ matrix can be easily obtained. Then one can also look for poles of the scattering amplitude $T_{ij}$ on the complex plane of $\sqrt{s}$. The pole, $Z_R$, on the second Riemann sheet could be associated with the $\Xi^*(1690)$ resonance. The real part of $Z_R$ is associated with the mass $M_{\Xi^*(1690)}$ of $\Xi^*(1690)$ resonance, and the imaginary part of $Z_R$ is associated with one half of its width $\Gamma_{\Xi^*(1690)}$.

\subsection{Contribution of $\Xi^*(1690)$ resonance as a Breit-Wigner resonance}

\begin{figure}[htbp]
    \includegraphics[scale = 1]{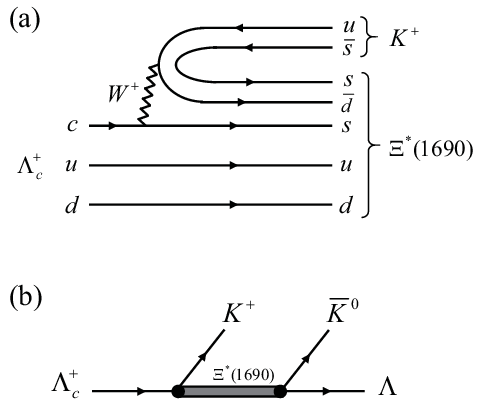}
\caption{(a) Quark level diagram for the $\Lambda_c^+ \to K^+\Xi^*(1690)$ decay and (b) hadron level diagram for the $\Lambda^+_c \to K^+\Xi^*(1690) \to K^+ \bar{K}^0 \Lambda$ decay.}
    \label{fig:xi1690bw}
\end{figure}

On the other hand, the decay process $\Lambda_c^+ \to \Xi^*(1690)K^+ \to \Lambda \bar{K}^0 K^+ $ can also proceed through $\Xi^*(1690)$ as a Breit-Wigner resonance and decaying into $\bar{K}^0 \Lambda$, which is shown in Fig.~\ref{fig:xi1690bw}. In this case, the $\Xi^*(1690)$ state is formed with $ud\bar{d}ss$ as shown in Fig.~\ref{fig:xi1690bw} (a), and it decays into $\bar{K}^0 \Lambda$ in the final state. The hadron level diagram for the decay of $\Lambda_c^+ \to \Xi^*(1690)K^+ \to \Lambda K^+\bar{K}^0$ is also shown in Fig.~\ref{fig:xi1690bw} (b), where the propagator of $\Xi^*(1690)$ resonance is parametrized as the Breit-Wigner form.

Then, the general decay amplitude for the process $\Lambda_c^+ \to \Xi^*(1690)^0 K^+ \to  \bar{K}^0 \Lambda K^+$ can be expressed as:
\begin{equation}
\mathcal{M}_e =  \frac{\Gamma_{\Xi^*(1690)}}{2}\frac{V_3}{M_{\bar{K}^0\Lambda}-M_{\Xi^*(1690)}+i\frac{\Gamma_{\Xi^*(1690)}}{2}},
\end{equation}
where $M_{\bar{K}^0\Lambda}$ is the invariant mass of the $\bar{K}^0 \Lambda$ system, and we take $M_{\Xi^*(1690)} = 1690$ MeV and $\Gamma_{\Xi^*(1690)} = 20$ MeV for the Breit-Wigner mass and width for $\Xi^*(1690)$ resonance, respectively, which are quoted in the PDG~\cite{ParticleDataGroup:2020ssz}. The model free parameter $V_3$ will be determined by the experimental data.

Before going further, we emphasize that the $ud\bar{d}ss$ component of $\Xi^*(1690)$ resonance cannot be guaranteed from the decay process shown in Fig.~\ref{fig:xi1690bw}. Indeed, the $\Xi^*(1690)$ resonance can also be produced from the $W$-exchange diagram~\cite{CLEO:1993fhs,Belle:2001hyr,BESIII:2018cvs,Gronau:2018vei}, where $cd$ transitions firstly into $su$ with the weak interaction, and then with a $s\bar{s}$ pair from the vacuum, the $u\bar{s}$ forms the $K^+$, while the $\Xi^*(1690)^0$ is constructed from the $uss$ cluster and then it decays into $\bar{K}^0 \Lambda$. Nevertheless, this kind of contributions can be absorbed into the model parameter $V_3$.

\subsection{Invariant mass distributions of the $\Lambda^+_c \to \Lambda K^+ \bar{K}^0$ decay}

With the ingredients obtained in the previous sections, we can write the double invariant mass distributions for the $\Lambda_c^+ \to K^+ \bar{K}^0 \Lambda$ decay as
\begin{equation}
     \frac{d^2\Gamma}{dM_{K^+\bar{K}^0} dM_{\bar{K}^0 \Lambda}} = \frac{M_\Lambda M_{K^+\bar{K}^0}M_{\bar{K}^0 \Lambda}}{16\pi^3M_{\Lambda_c^+}^2}|\mathcal{M}|^2, \label{eq:dgammadm12dm23}
\end{equation}
where ${\cal M}$ is the total decay amplitude. We will take two models for ${\cal M}$:
\begin{eqnarray}
\mathcal{M}^{\rm I}  &=& \mathcal{M}_a+\mathcal{M}_b+ \mathcal{M}_d + \mathcal{M}_c , \label{eq:totalMI}\\
\mathcal{M}^{\rm II} &=& \mathcal{M}_a + \mathcal{M}_b + \mathcal{M}_d + e^{i\theta}\mathcal{M}_e,
\label{eq:totalMII}
\end{eqnarray}
where a relative phase $\theta$ is added for Model II, and it is a free parameter. In addition, another relative phase, $\phi$, between $V_1$ and $V_2$ is also taken into account. We will replace $V_2$ with $e^{i\phi } V_2$ in the following fitting process. Furthermore, it is worth to mention that $V_1$, $V_2$ and $V_3$ have dimension ${\rm MeV}^{-1}$.

Then the invariant $\bar{K}^0\Lambda$ and $K^+ \bar{K}^0$ mass distributions can be obtained by integrating over the other invariant mass in Eq.~\eqref{eq:dgammadm12dm23}. For a given value of $M_{\bar{K}^0\Lambda}$, the invariant mass $M_{K^+\bar{K}^0}$ satisfies the following relation
\begin{equation}
\begin{aligned}
    (M_{K^+\bar{K}^0}^{\text{min}})^2&=(E_{K^+}+E_{\bar{K}^0})^2 \\
    &-(\sqrt{E_{K^+}^2-m_{K^+}^2}+\sqrt{E_{\bar{K}^0}^2-m_{\bar{K}^0}^2})^2, \\
    (M_{K^+\bar{K}^0}^{\text{max}})^2&=(E_{K^+}+E_{\bar{K}^0})^2  \\
    &-(\sqrt{E_{K^+}^2-m_{K^+}^2}-\sqrt{E_{\bar{K}^0}^2-m_{\bar{K}^0}^2})^2, 
\end{aligned} 
\end{equation}
where $E_{K^+}$ and $E_{\bar{K}^0}$ are the particle energies in the $\bar{K}^0\Lambda$ rest frame, which can be expressed explicitly
\begin{equation}
\begin{aligned}
    E_{K^+}&=\frac{M_{\Lambda_c^+}^2-M_{\bar{K}^0\Lambda}^2-M_{K^+}^2}{2M_{\bar{K}^0\Lambda}},\\
    E_{\bar{K}^0}&=\frac{M_{\bar{K}^0\Lambda}^2+M_{\bar{K}^0}^2-M_{\Lambda}^2}{2M_{\bar{K}^0\Lambda}}.
\end{aligned} 
\end{equation}

Note that the invariant $K^+\Lambda$ mass distribution can be obtained by substituting $M_{K^+\bar{K}^0}$ with $M_{K^+\Lambda}$.

\section{Results and Discussion} \label{sec:Results}

With the above formalism, we perform the $\chi^2$ fits to the experimental data on the invariant mass distributions of the process $\Lambda^+_c \to \Lambda K^+ \bar{K}^0$. There are a total of 79 data points. For Model I, there are five free parameters: $V_1$, $V_2$, $a_{\bar{K}\Sigma}$, $a_{\eta\Xi}$, and $\phi$. For both Model II, there are also five free parameters: $V_1$, $V_2$, $V_3$, $\phi$, and $\theta$. Note that we have fixed $a_{\pi\Xi} = a_{\bar{K}\Lambda}=-2$ as their natural value, and $a_{\bar{K}\Sigma}$ and $a_{\eta \Xi}$ are free parameters in this work. It is found that the pole position of $\Xi^*(1690)$ state is not sensitive to the values of $a_{\pi \Xi}$ and $a_{\bar{K}\Lambda}$, as found in Refs.~\cite{Ramos:2002xh,Sekihara:2015qqa}.

The fitted parameters and the corresponding $\chi^2/{\rm dof}$ are listed in table~\ref{table:pa-sets}. One can find that both of the two fits in table~\ref{table:pa-sets} show reasonably small $\chi^2/{\rm dof}$. With the fitted parameters $a_{\bar{K}\Sigma}$ and $a_{\eta\Xi}$, the pole position of $\Xi^*(1690)$ state is $M_R = 1685.83+i2.76$ MeV, and the corresponding mass and width are: $M_{\Xi^*(1690)} = 1685.83$ MeV, $\Gamma_{\Xi^*(1690)} = 5.52$ MeV. The obtained mass of $\Xi^*(1690)$ state is sightly below the mass threshold (1689 MeV) of the $\bar{K}\Sigma$ channel, and close to the averaged value quoted in the PDG. However, the obtained width is rather narrow, which is a common conclusion of the chiral unitary approach~\cite{Sekihara:2015qqa,Khemchandani:2016ftn}.

\begin{table}
\caption{Values of some of the parameters used or determined in this work. The values of $a_{\pi\Xi}$ and $a_{\bar{K}\Lambda}$ in Model~$\mathrm{I}$ are fixed as the natural value $-2$. The values of the mass and width for the $\Xi^*(1690)$ state are obtained from the fitted parameters of Model I. The mass and width of $\Xi^*(1690)$ resonance, for Model II, are taken as $M_{\Xi^*} = 1690$ MeV and $\Gamma_{\Xi^*} = 20$ MeV as quoted in PDG~\cite{ParticleDataGroup:2020ssz}.}
\label{table:pa-sets}
\setlength{\tabcolsep}{9pt}
\renewcommand{\arraystretch}{1.2}
\centering
\begin{tabular}{c|c|c} 
\hline                   
                     & Model~$\mathrm{I}$    & Model~$\mathrm{II}$  \\ 
\hline
$V_1$ (${\rm MeV}^{-1}$)               & 3.60$\pm$0.55  & 0.45$\pm$0.83  \\
$V_2$  (${\rm MeV}^{-1}$)               & -0.89$\pm$0.67  & -4.61$\pm$1.16  \\
$V_3$  (${\rm MeV}^{-1}$)               & $-$ &     1.81$\pm$0.18   \\
$a_{\bar{K}\Sigma}$  &  -1.99$\pm$0.09 & $-$   \\
$a_{\eta\Xi}$        &   -3.53$\pm$0.29    & $-$       \\
$\phi$               &  0.60$\pm$0.45 &3.20$\pm$1.16 \\
$\theta$             & $-$ &0.67$\pm$1.02 \\
$\chi^2/{\rm dof}$         & 1.51 & 1.56    \\
$M_{\Xi^*(1690)}$ (MeV)  & $1685.83$ (output)  & $1690$ (input)\\
$\Gamma_{\Xi^*(1690)}$ (MeV) & $5.52$ (output) &  $20$ (input)\\
\hline
\end{tabular}
\end{table}

\begin{figure}[htbp]
    \includegraphics[scale=0.35]{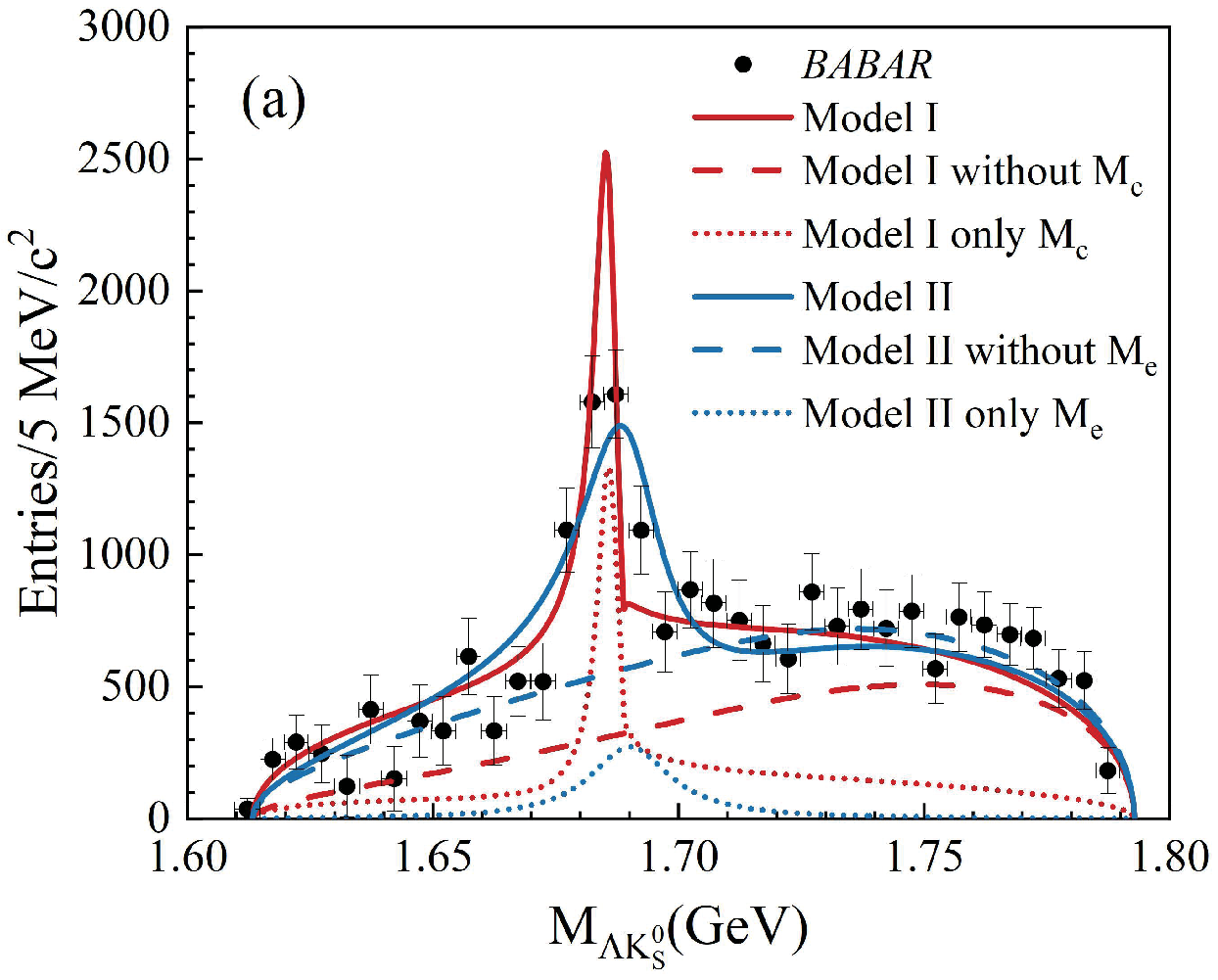}
    \includegraphics[scale=0.35]{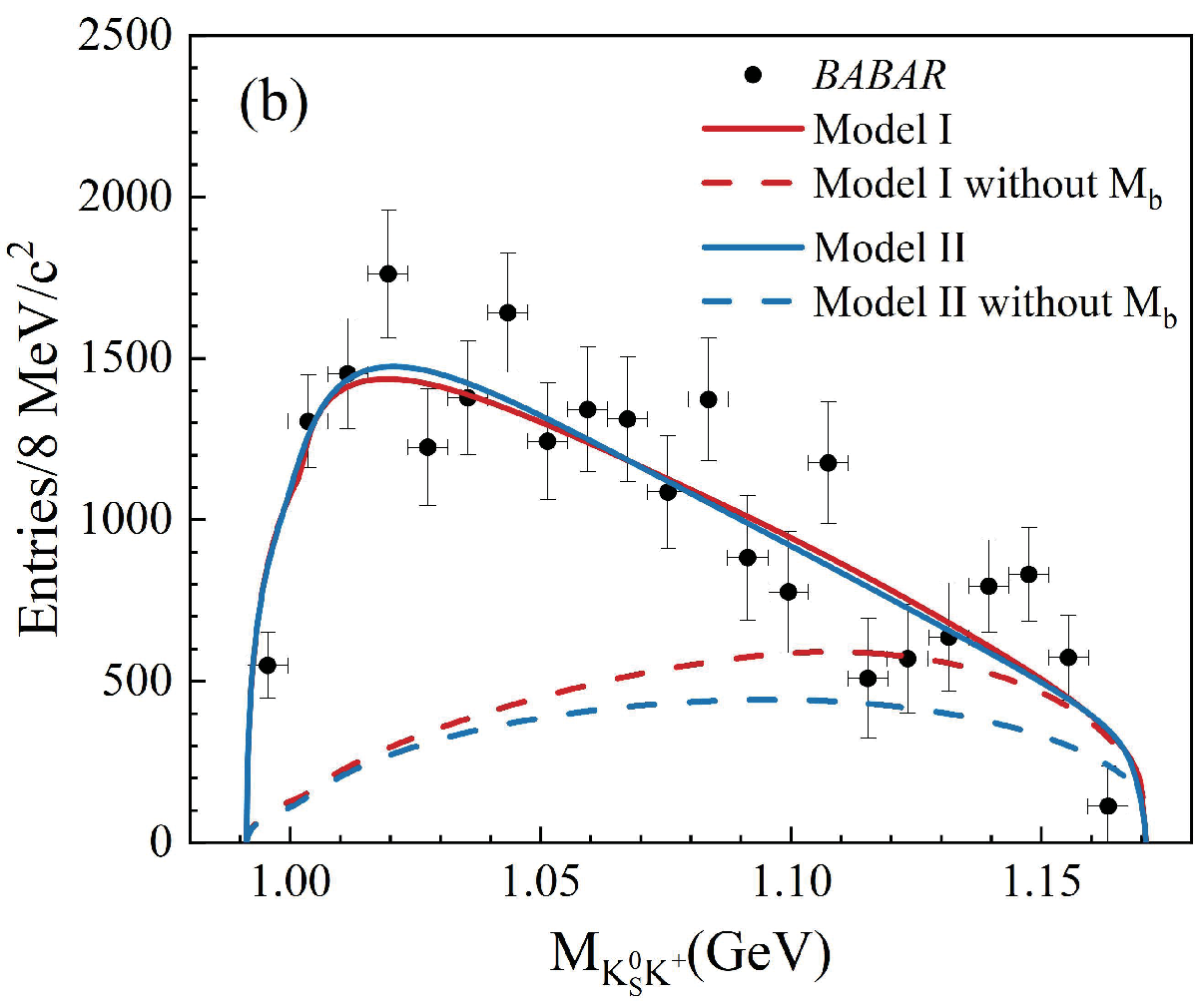}
    \includegraphics[scale=0.35]{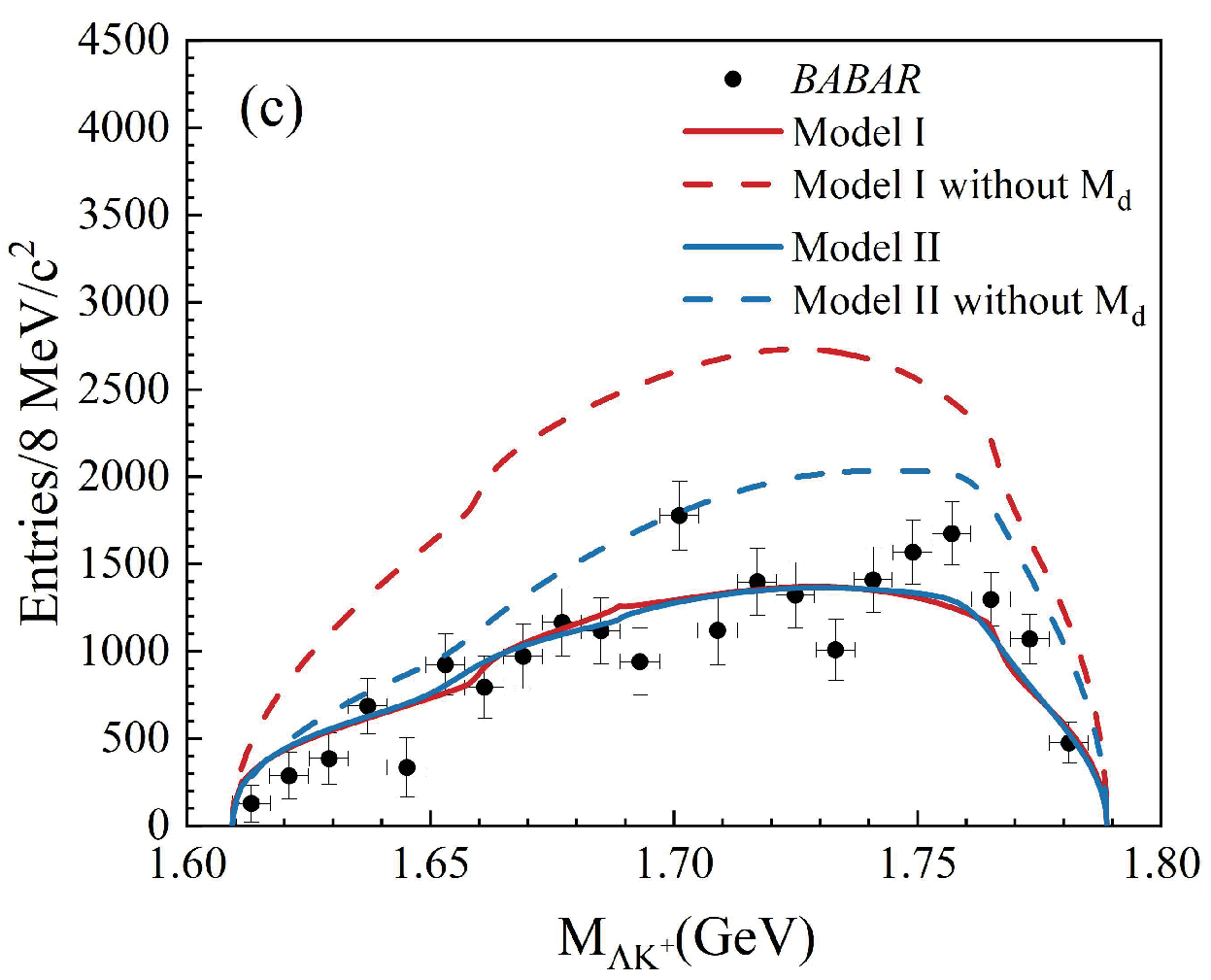}
\caption{(Color online) The invariant mass distributions for the $\Lambda_c^+ \to \Lambda K^+ K^0_S$ reaction: (a) $\Lambda K^0_S$ invariant mass distributions; (b) $K^+K^0_S$ invariant mass distribution; (c) $\Lambda K^+$ invariant mass distribution.}     \label{fig:fittedModelIandII}
\end{figure}

The fitted invariant mass distributions~\footnote{To compare with the experimental measurements, we take $\left|K^0\right\rangle=\frac{1}{\sqrt{2}} (\left|K^0_S\right\rangle+\left|K^0_L\right\rangle)$ and $\left|\bar{K}^0\right\rangle = \frac{1}{\sqrt{2}} (\left|K^0_S\right\rangle-\left|K^0_L\right\rangle)$, where we have ignored the small effect of $CP$ violation.} of the $\Lambda^+_c \to \Lambda K^+ \bar{K}^0$ reaction are shown in Fig.~\ref{fig:fittedModelIandII} for Model I and II. It is found that both Model I and II can describe these experimental measurements fairly well. In Fig.~\ref{fig:fittedModelIandII} (a), the peak of $\Xi^*(1690)$ resonance in the $\Lambda K^0_S$ invariant mass distributions is well reproduced.~\footnote{It is worth pointing out that we have added an extra factor $5/8$ to the theoretical results for the $\bar{K}^0\Lambda$ invariant mass distributions, since the experimental events were accumulated for a bin width of $5$ MeV, while the events for the invariant $K^+K^0_S$ and $K^+\Lambda$ mass distributions are obtained in a bin width of $8$ MeV.} And, it is clearly seen that the line shapes of $\Xi^*(1690)$ resonance are much different for Model I and Model II. The peak produced from the Model I is higher but narrower, and there is a sharp decrease around the mass threshold of the $\bar{K}\Sigma$ channel. It is expected that more precise data, around the energy of $M_{\bar{K}^0\Lambda} = 1.69$ GeV, could be used to clarify this issue. Furthermore, one can see that without the contributions of $\Xi^*(1690)$ (${\cal M}_c$ for Model I; ${\cal M}_e$ for Model II), the experimental data cannot be described.

In Figs.~\ref{fig:fittedModelIandII} (b) and (c), we show the $K^0_S K^+$ and $\Lambda K^+$ invariant mass distributions, respectively. It is found that the theoretical results of Model I and II are very similar. This may indicate that the contributions of $\Xi^*(1690)$ resonance, from Model I and II, to the $K^0_S K^+$ and $\Lambda K^+$ invariant mass distributions also are similar, though they give very different line shapes for $\Xi^*(1690)$ state. Once again, one can see that the contributions from the $K\bar{K}$ or $K\Lambda$ final state interactions in $S$-wave, as shown in Fig.~\ref{fig:final-s-i}, are crucial to reproduce the experimental measurements.

Next, we study the branching fractions of the $\Lambda^+_c \to K^+ \Xi^*(1690) \to K^+ \bar{K}^0 \Lambda$ decay to see the effect of the $\Xi^*(1690)$ resonance in the process of $\Lambda^+_c \to \Lambda K^+ \bar{K}^0$. With the full amplitude of Eqs.~\eqref{eq:totalMI} and \eqref{eq:totalMII} and the fitted parameters shown in table~\ref{table:pa-sets}, the numerical result for the branching fraction of the production of $\Xi^*(1690)$ resonance can be calculated as follows,
\begin{equation}
\frac{\mathcal{B}(\Lambda_c^+ \to K^+ \Xi^*(1690) \to K^+ \bar{K}^0 \Lambda)}{\mathcal{B}(\Lambda_c^+ \to \Lambda K^+\bar{K}^0)} = \begin{cases} 0.28 \\0.07 \end{cases},
    \label{eq:branch-lam-K0bar}
\end{equation}
for Model I and II, respectively. The above result of Model I is in good agreement with the experimental data of Belle collaboration: $0.26 \pm 0.08 \pm 0.03$~\cite{Belle:2001hyr}, while the result of the Model II is much smaller. We hope that the calculations here can be checked by future precise experiments.

\section{Summary} \label{sec:Summary}

In this work, we have performed an analysis of the $\Lambda_c^+\to\Lambda K^+\bar{K}^0$ decay via two cabibbo-favored mechanism. By considering the hadronization process and the final state interaction in the pseudoscalar-baryon and pseudoscalar-pseudoscalar channel with chiral unitary approach, the role of $\Xi^*(1690)$, $a_0(980)$, and $N^*(1535)$ resonances are investigated. Taking into account the contributions of these three above resonances, we have calculated the $K^+K_S^0$, $\Lambda K_S^0$ and $\Lambda K^+$ invariant mass distributions. Up to five free parameters, we have done $\chi^2$-fit to the experimental measurements. It was found that the reaction proposed here can describe the experimental data of $BABAR$ collaboration~\cite{BaBar:2006tck}. Especially, the clear signal for the $\Xi^*(1690)$ state can be well reproduced in the $\Lambda K^0_S$ invariant mass spectrum.

Within the fitted model parameters, we have further calculated the ratio of branching fractions $\mathcal{B}(\Lambda_c^+ \to K^+ \Xi^*(1690) \to K^+ \bar{K}^0 \Lambda)/\mathcal{B}(\Lambda_c^+ \to \Lambda K^+\bar{K}^0)$ to see the contribution of $\Xi^*(1690)$ resonance in the $\Lambda^+_c \to \Lambda K^+\bar{K}^0$ decay. The obtained value 0.28 for Model I, where the $\Xi^*(1690)$ is dynamically generated from the meson-baryon final state interactions, is consistent with the experimental value by Belle collaboration~\cite{Belle:2001hyr}. It is expected that the $\Xi^*(1690)$ state can be analyzed from the precise measurements of the $\Lambda^+_c \to K^+\bar{K}^0\Lambda$ decay by BESIII, BelleII, and LHCb collaborations in the future~\cite{BESIII:2023rky}.

\section*{Acknowledgments}

We would like to thank Profs. Wei-Hong Liang and Eulogio Oset for useful discussions. This work is partly supported by the National Natural Science Foundation of China under Grant No. 12075288. It is also supported by the Youth Innovation Promotion Association CAS.


\bibliography{reference.bib}

\begin{thebibliography}{62}%
\makeatletter
\providecommand \@ifxundefined [1]{%
 \@ifx{#1\undefined}
}%
\providecommand \@ifnum [1]{%
 \ifnum #1\expandafter \@firstoftwo
 \else \expandafter \@secondoftwo
 \fi
}%
\providecommand \@ifx [1]{%
 \ifx #1\expandafter \@firstoftwo
 \else \expandafter \@secondoftwo
 \fi
}%
\providecommand \natexlab [1]{#1}%
\providecommand \enquote  [1]{``#1''}%
\providecommand \bibnamefont  [1]{#1}%
\providecommand \bibfnamefont [1]{#1}%
\providecommand \citenamefont [1]{#1}%
\providecommand \href@noop [0]{\@secondoftwo}%
\providecommand \href [0]{\begingroup \@sanitize@url \@href}%
\providecommand \@href[1]{\@@startlink{#1}\@@href}%
\providecommand \@@href[1]{\endgroup#1\@@endlink}%
\providecommand \@sanitize@url [0]{\catcode `\\12\catcode `\$12\catcode
  `\&12\catcode `\#12\catcode `\^12\catcode `\_12\catcode `\%12\relax}%
\providecommand \@@startlink[1]{}%
\providecommand \@@endlink[0]{}%
\providecommand \url  [0]{\begingroup\@sanitize@url \@url }%
\providecommand \@url [1]{\endgroup\@href {#1}{\urlprefix }}%
\providecommand \urlprefix  [0]{URL }%
\providecommand \Eprint [0]{\href }%
\providecommand \doibase [0]{https://doi.org/}%
\providecommand \selectlanguage [0]{\@gobble}%
\providecommand \bibinfo  [0]{\@secondoftwo}%
\providecommand \bibfield  [0]{\@secondoftwo}%
\providecommand \translation [1]{[#1]}%
\providecommand \BibitemOpen [0]{}%
\providecommand \bibitemStop [0]{}%
\providecommand \bibitemNoStop [0]{.\EOS\space}%
\providecommand \EOS [0]{\spacefactor3000\relax}%
\providecommand \BibitemShut  [1]{\csname bibitem#1\endcsname}%
\let\auto@bib@innerbib\@empty
\bibitem [{\citenamefont {Garcia-Recio}\ \emph {et~al.}(2004)\citenamefont
  {Garcia-Recio}, \citenamefont {Lutz},\ and\ \citenamefont
  {Nieves}}]{Garcia-Recio:2003ejq}%
  \BibitemOpen
  \bibfield  {author} {\bibinfo {author} {\bibfnamefont {C.}~\bibnamefont
  {Garcia-Recio}}, \bibinfo {author} {\bibfnamefont {M.~F.~M.}\ \bibnamefont
  {Lutz}},\ and\ \bibinfo {author} {\bibfnamefont {J.}~\bibnamefont {Nieves}},\
  }\bibfield  {title} {\bibinfo {title} {{Quark mass dependence of s wave
  baryon resonances}},\ }\href {https://doi.org/10.1016/j.physletb.2003.11.073}
  {\bibfield  {journal} {\bibinfo  {journal} {Phys. Lett. B}\ }\textbf
  {\bibinfo {volume} {582}},\ \bibinfo {pages} {49} (\bibinfo {year} {2004})},\
  \Eprint {https://arxiv.org/abs/nucl-th/0305100} {arXiv:nucl-th/0305100}
  \BibitemShut {NoStop}%
\bibitem [{\citenamefont {Gamermann}\ \emph {et~al.}(2011)\citenamefont
  {Gamermann}, \citenamefont {Garcia-Recio}, \citenamefont {Nieves},\ and\
  \citenamefont {Salcedo}}]{Gamermann:2011mq}%
  \BibitemOpen
  \bibfield  {author} {\bibinfo {author} {\bibfnamefont {D.}~\bibnamefont
  {Gamermann}}, \bibinfo {author} {\bibfnamefont {C.}~\bibnamefont
  {Garcia-Recio}}, \bibinfo {author} {\bibfnamefont {J.}~\bibnamefont
  {Nieves}},\ and\ \bibinfo {author} {\bibfnamefont {L.~L.}\ \bibnamefont
  {Salcedo}},\ }\bibfield  {title} {\bibinfo {title} {{Odd Parity Light Baryon
  Resonances}},\ }\href {https://doi.org/10.1103/PhysRevD.84.056017} {\bibfield
   {journal} {\bibinfo  {journal} {Phys. Rev. D}\ }\textbf {\bibinfo {volume}
  {84}},\ \bibinfo {pages} {056017} (\bibinfo {year} {2011})},\ \Eprint
  {https://arxiv.org/abs/1104.2737} {arXiv:1104.2737 [hep-ph]} \BibitemShut
  {NoStop}%
\bibitem [{\citenamefont {Pervin}\ and\ \citenamefont
  {Roberts}(2008)}]{Pervin:2007wa}%
  \BibitemOpen
  \bibfield  {author} {\bibinfo {author} {\bibfnamefont {M.}~\bibnamefont
  {Pervin}}\ and\ \bibinfo {author} {\bibfnamefont {W.}~\bibnamefont
  {Roberts}},\ }\bibfield  {title} {\bibinfo {title} {{Strangeness -2 and -3
  baryons in a constituent quark model}},\ }\href
  {https://doi.org/10.1103/PhysRevC.77.025202} {\bibfield  {journal} {\bibinfo
  {journal} {Phys. Rev. C}\ }\textbf {\bibinfo {volume} {77}},\ \bibinfo
  {pages} {025202} (\bibinfo {year} {2008})},\ \Eprint
  {https://arxiv.org/abs/0709.4000} {arXiv:0709.4000 [nucl-th]} \BibitemShut
  {NoStop}%
\bibitem [{\citenamefont {Pavon~Valderrama}\ \emph {et~al.}(2012)\citenamefont
  {Pavon~Valderrama}, \citenamefont {Xie},\ and\ \citenamefont
  {Nieves}}]{PavonValderrama:2011gp}%
  \BibitemOpen
  \bibfield  {author} {\bibinfo {author} {\bibfnamefont {M.}~\bibnamefont
  {Pavon~Valderrama}}, \bibinfo {author} {\bibfnamefont {J.-J.}\ \bibnamefont
  {Xie}},\ and\ \bibinfo {author} {\bibfnamefont {J.}~\bibnamefont {Nieves}},\
  }\bibfield  {title} {\bibinfo {title} {{Are there three $\Xi(1950)$
  states?}},\ }\href {https://doi.org/10.1103/PhysRevD.85.017502} {\bibfield
  {journal} {\bibinfo  {journal} {Phys. Rev. D}\ }\textbf {\bibinfo {volume}
  {85}},\ \bibinfo {pages} {017502} (\bibinfo {year} {2012})},\ \Eprint
  {https://arxiv.org/abs/1111.2218} {arXiv:1111.2218 [hep-ph]} \BibitemShut
  {NoStop}%
\bibitem [{\citenamefont {Nishibuchi}\ and\ \citenamefont
  {Hyodo}(2022)}]{Nishibuchi:2022zfo}%
  \BibitemOpen
  \bibfield  {author} {\bibinfo {author} {\bibfnamefont {T.}~\bibnamefont
  {Nishibuchi}}\ and\ \bibinfo {author} {\bibfnamefont {T.}~\bibnamefont
  {Hyodo}},\ }\bibfield  {title} {\bibinfo {title} {{Nature of excited $\Xi$
  baryons with threshold effects}},\ }\href
  {https://doi.org/10.1051/epjconf/202227110002} {\bibfield  {journal}
  {\bibinfo  {journal} {EPJ Web Conf.}\ }\textbf {\bibinfo {volume} {271}},\
  \bibinfo {pages} {10002} (\bibinfo {year} {2022})},\ \Eprint
  {https://arxiv.org/abs/2208.14608} {arXiv:2208.14608 [hep-ph]} \BibitemShut
  {NoStop}%
\bibitem [{\citenamefont {Zyla}\ \emph {et~al.}(2020)\citenamefont {Zyla} \emph
  {et~al.}}]{ParticleDataGroup:2020ssz}%
  \BibitemOpen
  \bibfield  {author} {\bibinfo {author} {\bibfnamefont {P.~A.}\ \bibnamefont
  {Zyla}} \emph {et~al.} (\bibinfo {collaboration} {Particle Data Group}),\
  }\bibfield  {title} {\bibinfo {title} {{Review of Particle Physics}},\ }\href
  {https://doi.org/10.1093/ptep/ptaa104} {\bibfield  {journal} {\bibinfo
  {journal} {PTEP}\ }\textbf {\bibinfo {volume} {2020}},\ \bibinfo {pages}
  {083C01} (\bibinfo {year} {2020})}\BibitemShut {NoStop}%
\bibitem [{\citenamefont {Nagae}(2005)}]{Nagae:2005rj}%
  \BibitemOpen
  \bibfield  {author} {\bibinfo {author} {\bibfnamefont {T.}~\bibnamefont
  {Nagae}},\ }\bibfield  {title} {\bibinfo {title} {{The J-PARC project}},\
  }\href {https://doi.org/10.1063/1.1949488} {\bibfield  {journal} {\bibinfo
  {journal} {AIP Conf. Proc.}\ }\textbf {\bibinfo {volume} {773}},\ \bibinfo
  {pages} {6} (\bibinfo {year} {2005})}\BibitemShut {NoStop}%
\bibitem [{\citenamefont {Guo}\ \emph {et~al.}(2007)\citenamefont {Guo} \emph
  {et~al.}}]{Guo:2007dw}%
  \BibitemOpen
  \bibfield  {author} {\bibinfo {author} {\bibfnamefont {L.}~\bibnamefont
  {Guo}} \emph {et~al.},\ }\bibfield  {title} {\bibinfo {title} {{Cascade
  production in the reactions $\gamma p \to K^+ K^+ (X)$ and $\gamma p \to K^+
  K^+ \pi^- (X)$}},\ }\href {https://doi.org/10.1103/PhysRevC.76.025208}
  {\bibfield  {journal} {\bibinfo  {journal} {Phys. Rev. C}\ }\textbf {\bibinfo
  {volume} {76}},\ \bibinfo {pages} {025208} (\bibinfo {year} {2007})},\
  \Eprint {https://arxiv.org/abs/nucl-ex/0702027} {arXiv:nucl-ex/0702027}
  \BibitemShut {NoStop}%
\bibitem [{\citenamefont {Schumacher}(2010)}]{Schumacher:2010zz}%
  \BibitemOpen
  \bibfield  {author} {\bibinfo {author} {\bibfnamefont {R.}~\bibnamefont
  {Schumacher}} (\bibinfo {collaboration} {CLAS}),\ }\bibfield  {title}
  {\bibinfo {title} {{Strangeness physics with CLAS at Jefferson Lab}},\ }\href
  {https://doi.org/10.1063/1.3483304} {\bibfield  {journal} {\bibinfo
  {journal} {AIP Conf. Proc.}\ }\textbf {\bibinfo {volume} {1257}},\ \bibinfo
  {pages} {100} (\bibinfo {year} {2010})}\BibitemShut {NoStop}%
\bibitem [{\citenamefont {Lutz}\ \emph {et~al.}(2009)\citenamefont {Lutz} \emph
  {et~al.}}]{PANDA:2009yku}%
  \BibitemOpen
  \bibfield  {author} {\bibinfo {author} {\bibfnamefont {M.~F.~M.}\
  \bibnamefont {Lutz}} \emph {et~al.} (\bibinfo {collaboration} {PANDA}),\
  }\bibfield  {title} {\bibinfo {title} {{Physics Performance Report for PANDA:
  Strong Interaction Studies with Antiprotons}},\ }\href@noop {} {\  (\bibinfo
  {year} {2009})},\ \Eprint {https://arxiv.org/abs/0903.3905} {arXiv:0903.3905
  [hep-ex]} \BibitemShut {NoStop}%
\bibitem [{\citenamefont {Ahn}\ and\ \citenamefont {Nam}(2018)}]{Ahn:2018hbc}%
  \BibitemOpen
  \bibfield  {author} {\bibinfo {author} {\bibfnamefont {J.~K.}\ \bibnamefont
  {Ahn}}\ and\ \bibinfo {author} {\bibfnamefont {S.-i.}\ \bibnamefont {Nam}},\
  }\bibfield  {title} {\bibinfo {title} {{$\Xi(1690)^-$ production in the
  $K^-p\to K^+K^-\Lambda$ reaction process near threshold}},\ }\href
  {https://doi.org/10.1103/PhysRevD.98.114012} {\bibfield  {journal} {\bibinfo
  {journal} {Phys. Rev. D}\ }\textbf {\bibinfo {volume} {98}},\ \bibinfo
  {pages} {114012} (\bibinfo {year} {2018})},\ \Eprint
  {https://arxiv.org/abs/1809.07765} {arXiv:1809.07765 [hep-ph]} \BibitemShut
  {NoStop}%
\bibitem [{\citenamefont {Dionisi}\ \emph {et~al.}(1978)\citenamefont {Dionisi}
  \emph {et~al.}}]{Amsterdam-CERN-Nijmegen-Oxford:1978chc}%
  \BibitemOpen
  \bibfield  {author} {\bibinfo {author} {\bibfnamefont {C.}~\bibnamefont
  {Dionisi}} \emph {et~al.} (\bibinfo {collaboration}
  {Amsterdam-CERN-Nijmegen-Oxford}),\ }\bibfield  {title} {\bibinfo {title}
  {{An Enhancement at the $\Sigma \bar{K}$ Threshold 1680-{MeV} Observed in
  $K^- p$ Reactions at 4.2-{GeV}/$c$}},\ }\href
  {https://doi.org/10.1016/0370-2693(78)90329-5} {\bibfield  {journal}
  {\bibinfo  {journal} {Phys. Lett. B}\ }\textbf {\bibinfo {volume} {80}},\
  \bibinfo {pages} {145} (\bibinfo {year} {1978})}\BibitemShut {NoStop}%
\bibitem [{\citenamefont {Adamovich}\ \emph {et~al.}(1998)\citenamefont
  {Adamovich} \emph {et~al.}}]{WA89:1997vxx}%
  \BibitemOpen
  \bibfield  {author} {\bibinfo {author} {\bibfnamefont {M.~I.}\ \bibnamefont
  {Adamovich}} \emph {et~al.} (\bibinfo {collaboration} {WA89}),\ }\bibfield
  {title} {\bibinfo {title} {{First observation of the $\Xi^- \pi^+$ decay mode
  of the $\Xi^0(1690)$ hyperon}},\ }\href
  {https://doi.org/10.1007/s100520050304} {\bibfield  {journal} {\bibinfo
  {journal} {Eur. Phys. J. C}\ }\textbf {\bibinfo {volume} {5}},\ \bibinfo
  {pages} {621} (\bibinfo {year} {1998})},\ \Eprint
  {https://arxiv.org/abs/hep-ex/9710024} {arXiv:hep-ex/9710024} \BibitemShut
  {NoStop}%
\bibitem [{\citenamefont {Aubert}\ \emph {et~al.}(2008)\citenamefont {Aubert}
  \emph {et~al.}}]{BaBar:2008myc}%
  \BibitemOpen
  \bibfield  {author} {\bibinfo {author} {\bibfnamefont {B.}~\bibnamefont
  {Aubert}} \emph {et~al.} (\bibinfo {collaboration} {BaBar}),\ }\bibfield
  {title} {\bibinfo {title} {{Measurement of the Spin of the $\Xi(1530)$
  Resonance}},\ }\href {https://doi.org/10.1103/PhysRevD.78.034008} {\bibfield
  {journal} {\bibinfo  {journal} {Phys. Rev. D}\ }\textbf {\bibinfo {volume}
  {78}},\ \bibinfo {pages} {034008} (\bibinfo {year} {2008})},\ \Eprint
  {https://arxiv.org/abs/0803.1863} {arXiv:0803.1863 [hep-ex]} \BibitemShut
  {NoStop}%
\bibitem [{\citenamefont {Abe}\ \emph {et~al.}(2002)\citenamefont {Abe} \emph
  {et~al.}}]{Belle:2001hyr}%
  \BibitemOpen
  \bibfield  {author} {\bibinfo {author} {\bibfnamefont {K.}~\bibnamefont
  {Abe}} \emph {et~al.} (\bibinfo {collaboration} {Belle}),\ }\bibfield
  {title} {\bibinfo {title} {{Observation of Cabibbo suppressed and W exchange
  $\Lambda+_c$ baryon decays}},\ }\href
  {https://doi.org/10.1016/S0370-2693(01)01373-9} {\bibfield  {journal}
  {\bibinfo  {journal} {Phys. Lett. B}\ }\textbf {\bibinfo {volume} {524}},\
  \bibinfo {pages} {33} (\bibinfo {year} {2002})},\ \Eprint
  {https://arxiv.org/abs/hep-ex/0111032} {arXiv:hep-ex/0111032} \BibitemShut
  {NoStop}%
\bibitem [{\citenamefont {Aubert}\ \emph {et~al.}(2006)\citenamefont {Aubert}
  \emph {et~al.}}]{BaBar:2006tck}%
  \BibitemOpen
  \bibfield  {author} {\bibinfo {author} {\bibfnamefont {B.}~\bibnamefont
  {Aubert}} \emph {et~al.} (\bibinfo {collaboration} {BaBar}),\ }\bibfield
  {title} {\bibinfo {title} {{Measurement of the Mass and Width and Study of
  the Spin of the $\Xi(1690) $ 0 Resonance from $\Lambda^+_{c} \to \Lambda
  \bar{K}^0 K^{+}$ Decay at Babar}},\ }in\ \href@noop {} {\emph {\bibinfo
  {booktitle} {{33rd International Conference on High Energy Physics}}}}\
  (\bibinfo {year} {2006})\ \Eprint {https://arxiv.org/abs/hep-ex/0607043}
  {arXiv:hep-ex/0607043} \BibitemShut {NoStop}%
\bibitem [{\citenamefont {Biagi}\ \emph {et~al.}(1981)\citenamefont {Biagi}
  \emph {et~al.}}]{Biagi:1981cu}%
  \BibitemOpen
  \bibfield  {author} {\bibinfo {author} {\bibfnamefont {S.~F.}\ \bibnamefont
  {Biagi}} \emph {et~al.},\ }\bibfield  {title} {\bibinfo {title} {{Production
  of Hyperons and Hyperon Resonances in $\Xi^- N$ Interactions at 102-{GeV}/$c$
  and 135-{GeV}/$c$}},\ }\href {https://doi.org/10.1007/BF01548765} {\bibfield
  {journal} {\bibinfo  {journal} {Z. Phys. C}\ }\textbf {\bibinfo {volume}
  {9}},\ \bibinfo {pages} {305} (\bibinfo {year} {1981})}\BibitemShut {NoStop}%
\bibitem [{\citenamefont {Biagi}\ \emph {et~al.}(1987)\citenamefont {Biagi}
  \emph {et~al.}}]{Biagi:1986zj}%
  \BibitemOpen
  \bibfield  {author} {\bibinfo {author} {\bibfnamefont {S.~F.}\ \bibnamefont
  {Biagi}} \emph {et~al.},\ }\bibfield  {title} {\bibinfo {title} {{$\Xi^*$
  Resonances in $\Xi^-$ Be Interactions. 1. Diffractive Production in the
  $\Lambda K^-$ and $\Xi^- \pi^+ \pi^-$ Channels}},\ }\href
  {https://doi.org/10.1007/BF01561109} {\bibfield  {journal} {\bibinfo
  {journal} {Z. Phys. C}\ }\textbf {\bibinfo {volume} {34}},\ \bibinfo {pages}
  {15} (\bibinfo {year} {1987})}\BibitemShut {NoStop}%
\bibitem [{\citenamefont {Sumihama}\ \emph {et~al.}(2019)\citenamefont
  {Sumihama} \emph {et~al.}}]{Belle:2018lws}%
  \BibitemOpen
  \bibfield  {author} {\bibinfo {author} {\bibfnamefont {M.}~\bibnamefont
  {Sumihama}} \emph {et~al.} (\bibinfo {collaboration} {Belle}),\ }\bibfield
  {title} {\bibinfo {title} {{Observation of $\Xi(1620)^0$ and evidence for
  $\Xi(1690)^0$ in $\Xi_c^+ \rightarrow \Xi^-\pi^+\pi^+$ decays}},\ }\href
  {https://doi.org/10.1103/PhysRevLett.122.072501} {\bibfield  {journal}
  {\bibinfo  {journal} {Phys. Rev. Lett.}\ }\textbf {\bibinfo {volume} {122}},\
  \bibinfo {pages} {072501} (\bibinfo {year} {2019})},\ \Eprint
  {https://arxiv.org/abs/1810.06181} {arXiv:1810.06181 [hep-ex]} \BibitemShut
  {NoStop}%
\bibitem [{\citenamefont {Chao}\ \emph {et~al.}(1981)\citenamefont {Chao},
  \citenamefont {Isgur},\ and\ \citenamefont {Karl}}]{Chao:1980em}%
  \BibitemOpen
  \bibfield  {author} {\bibinfo {author} {\bibfnamefont {K.-T.}\ \bibnamefont
  {Chao}}, \bibinfo {author} {\bibfnamefont {N.}~\bibnamefont {Isgur}},\ and\
  \bibinfo {author} {\bibfnamefont {G.}~\bibnamefont {Karl}},\ }\bibfield
  {title} {\bibinfo {title} {{Strangeness -2 and -3 Baryons in a Quark Model
  With Chromodynamics}},\ }\href {https://doi.org/10.1103/PhysRevD.23.155}
  {\bibfield  {journal} {\bibinfo  {journal} {Phys. Rev. D}\ }\textbf {\bibinfo
  {volume} {23}},\ \bibinfo {pages} {155} (\bibinfo {year} {1981})}\BibitemShut
  {NoStop}%
\bibitem [{\citenamefont {Capstick}\ and\ \citenamefont
  {Isgur}(1986)}]{Capstick:1986ter}%
  \BibitemOpen
  \bibfield  {author} {\bibinfo {author} {\bibfnamefont {S.}~\bibnamefont
  {Capstick}}\ and\ \bibinfo {author} {\bibfnamefont {N.}~\bibnamefont
  {Isgur}},\ }\bibfield  {title} {\bibinfo {title} {{Baryons in a relativized
  quark model with chromodynamics}},\ }\href
  {https://doi.org/10.1103/physrevd.34.2809} {\bibfield  {journal} {\bibinfo
  {journal} {Phys. Rev. D}\ }\textbf {\bibinfo {volume} {34}},\ \bibinfo
  {pages} {2809} (\bibinfo {year} {1986})}\BibitemShut {NoStop}%
\bibitem [{\citenamefont {Xiao}\ and\ \citenamefont
  {Zhong}(2013)}]{Xiao:2013xi}%
  \BibitemOpen
  \bibfield  {author} {\bibinfo {author} {\bibfnamefont {L.-Y.}\ \bibnamefont
  {Xiao}}\ and\ \bibinfo {author} {\bibfnamefont {X.-H.}\ \bibnamefont
  {Zhong}},\ }\bibfield  {title} {\bibinfo {title} {{$\Xi$ baryon strong decays
  in a chiral quark model}},\ }\href
  {https://doi.org/10.1103/PhysRevD.87.094002} {\bibfield  {journal} {\bibinfo
  {journal} {Phys. Rev. D}\ }\textbf {\bibinfo {volume} {87}},\ \bibinfo
  {pages} {094002} (\bibinfo {year} {2013})},\ \Eprint
  {https://arxiv.org/abs/1302.0079} {arXiv:1302.0079 [hep-ph]} \BibitemShut
  {NoStop}%
\bibitem [{\citenamefont {Sekihara}(2015)}]{Sekihara:2015qqa}%
  \BibitemOpen
  \bibfield  {author} {\bibinfo {author} {\bibfnamefont {T.}~\bibnamefont
  {Sekihara}},\ }\bibfield  {title} {\bibinfo {title} {{$\Xi (1690)$ as a
  $\bar{K} \Sigma$ molecular state}},\ }\href
  {https://doi.org/10.1093/ptep/ptv129} {\bibfield  {journal} {\bibinfo
  {journal} {PTEP}\ }\textbf {\bibinfo {volume} {2015}},\ \bibinfo {pages}
  {091D01} (\bibinfo {year} {2015})},\ \Eprint
  {https://arxiv.org/abs/1505.02849} {arXiv:1505.02849 [hep-ph]} \BibitemShut
  {NoStop}%
\bibitem [{\citenamefont {Aliev}\ \emph {et~al.}(2018)\citenamefont {Aliev},
  \citenamefont {Azizi},\ and\ \citenamefont {Sundu}}]{Aliev:2018hre}%
  \BibitemOpen
  \bibfield  {author} {\bibinfo {author} {\bibfnamefont {T.~M.}\ \bibnamefont
  {Aliev}}, \bibinfo {author} {\bibfnamefont {K.}~\bibnamefont {Azizi}},\ and\
  \bibinfo {author} {\bibfnamefont {H.}~\bibnamefont {Sundu}},\ }\bibfield
  {title} {\bibinfo {title} {{Analysis of the structure of $\Xi (1690)$ through
  its decays}},\ }\href {https://doi.org/10.1140/epjc/s10052-018-5888-8}
  {\bibfield  {journal} {\bibinfo  {journal} {Eur. Phys. J. C}\ }\textbf
  {\bibinfo {volume} {78}},\ \bibinfo {pages} {396} (\bibinfo {year} {2018})},\
  \Eprint {https://arxiv.org/abs/1804.02656} {arXiv:1804.02656 [hep-ph]}
  \BibitemShut {NoStop}%
\bibitem [{\citenamefont {Khemchandani}\ \emph {et~al.}(2018)\citenamefont
  {Khemchandani}, \citenamefont {Mart\'\i{}nez~Torres}, \citenamefont {Hosaka},
  \citenamefont {Nagahiro}, \citenamefont {Navarra},\ and\ \citenamefont
  {Nielsen}}]{Khemchandani:2016ftn}%
  \BibitemOpen
  \bibfield  {author} {\bibinfo {author} {\bibfnamefont {K.~P.}\ \bibnamefont
  {Khemchandani}}, \bibinfo {author} {\bibfnamefont {A.}~\bibnamefont
  {Mart\'\i{}nez~Torres}}, \bibinfo {author} {\bibfnamefont {A.}~\bibnamefont
  {Hosaka}}, \bibinfo {author} {\bibfnamefont {H.}~\bibnamefont {Nagahiro}},
  \bibinfo {author} {\bibfnamefont {F.~S.}\ \bibnamefont {Navarra}},\ and\
  \bibinfo {author} {\bibfnamefont {M.}~\bibnamefont {Nielsen}},\ }\bibfield
  {title} {\bibinfo {title} {{Why $\Xi(1690)$ and $\Xi(2120)$ are so
  narrow?}},\ }\href {https://doi.org/10.1103/PhysRevD.97.034005} {\bibfield
  {journal} {\bibinfo  {journal} {Phys. Rev. D}\ }\textbf {\bibinfo {volume}
  {97}},\ \bibinfo {pages} {034005} (\bibinfo {year} {2018})},\ \Eprint
  {https://arxiv.org/abs/1608.07086} {arXiv:1608.07086 [nucl-th]} \BibitemShut
  {NoStop}%
\bibitem [{\citenamefont {Feijoo}\ \emph {et~al.}(2023)\citenamefont {Feijoo},
  \citenamefont {Valcarce},\ and\ \citenamefont {Magas}}]{Feijoo:2023wua}%
  \BibitemOpen
  \bibfield  {author} {\bibinfo {author} {\bibfnamefont {A.}~\bibnamefont
  {Feijoo}}, \bibinfo {author} {\bibfnamefont {V.}~\bibnamefont {Valcarce}},\
  and\ \bibinfo {author} {\bibfnamefont {V.~K.}\ \bibnamefont {Magas}},\
  }\bibfield  {title} {\bibinfo {title} {{The $\Xi(1620)$ and $\Xi(1690)$
  molecular states from $S=-2$ meson-baryon interaction up to next-to-leading
  order}},\ }\href@noop {} {\  (\bibinfo {year} {2023})},\ \Eprint
  {https://arxiv.org/abs/2303.01323} {arXiv:2303.01323 [hep-ph]} \BibitemShut
  {NoStop}%
\bibitem [{\citenamefont {Oset}\ \emph
  {et~al.}(2016{\natexlab{a}})\citenamefont {Oset} \emph
  {et~al.}}]{Oset:2016lyh}%
  \BibitemOpen
  \bibfield  {author} {\bibinfo {author} {\bibfnamefont {E.}~\bibnamefont
  {Oset}} \emph {et~al.},\ }\bibfield  {title} {\bibinfo {title} {{Weak decays
  of heavy hadrons into dynamically generated resonances}},\ }\href
  {https://doi.org/10.1142/S0218301316300010} {\bibfield  {journal} {\bibinfo
  {journal} {Int. J. Mod. Phys. E}\ }\textbf {\bibinfo {volume} {25}},\
  \bibinfo {pages} {1630001} (\bibinfo {year} {2016}{\natexlab{a}})},\ \Eprint
  {https://arxiv.org/abs/1601.03972} {arXiv:1601.03972 [hep-ph]} \BibitemShut
  {NoStop}%
\bibitem [{\citenamefont {Oset}\ \emph
  {et~al.}(2016{\natexlab{b}})\citenamefont {Oset} \emph
  {et~al.}}]{Oset:2016nvf}%
  \BibitemOpen
  \bibfield  {author} {\bibinfo {author} {\bibfnamefont {E.}~\bibnamefont
  {Oset}} \emph {et~al.},\ }\bibfield  {title} {\bibinfo {title} {{Study of
  reactions disclosing hidden charm pentaquarks with or without strangeness}},\
  }\href {https://doi.org/10.1016/j.nuclphysa.2016.04.038} {\bibfield
  {journal} {\bibinfo  {journal} {Nucl. Phys. A}\ }\textbf {\bibinfo {volume}
  {954}},\ \bibinfo {pages} {371} (\bibinfo {year}
  {2016}{\natexlab{b}})}\BibitemShut {NoStop}%
\bibitem [{\citenamefont {Yu}\ and\ \citenamefont {Hsiao}(2021)}]{Yu:2020vlt}%
  \BibitemOpen
  \bibfield  {author} {\bibinfo {author} {\bibfnamefont {Y.}~\bibnamefont
  {Yu}}\ and\ \bibinfo {author} {\bibfnamefont {Y.-K.}\ \bibnamefont {Hsiao}},\
  }\bibfield  {title} {\bibinfo {title} {{Cabibbo-favored
  $\ensuremath{\Lambda}_c^+\to\ensuremath{\Lambda}a_0(980)^+$ decay in the
  final state interaction}},\ }\href
  {https://doi.org/10.1016/j.physletb.2021.136586} {\bibfield  {journal}
  {\bibinfo  {journal} {Phys. Lett. B}\ }\textbf {\bibinfo {volume} {820}},\
  \bibinfo {pages} {136586} (\bibinfo {year} {2021})},\ \Eprint
  {https://arxiv.org/abs/2012.14575} {arXiv:2012.14575 [hep-ph]} \BibitemShut
  {NoStop}%
\bibitem [{\citenamefont {Ahn}\ \emph {et~al.}(2019)\citenamefont {Ahn},
  \citenamefont {Yang},\ and\ \citenamefont {Nam}}]{Ahn:2019rdr}%
  \BibitemOpen
  \bibfield  {author} {\bibinfo {author} {\bibfnamefont {J.~K.}\ \bibnamefont
  {Ahn}}, \bibinfo {author} {\bibfnamefont {S.}~\bibnamefont {Yang}},\ and\
  \bibinfo {author} {\bibfnamefont {S.-I.}\ \bibnamefont {Nam}},\ }\bibfield
  {title} {\bibinfo {title} {{Hyperon production in $\Lambda_c^+\to K^-p\pi^+$
  and $\Lambda_c^+\to K^0_Sp\pi^0 $}},\ }\href
  {https://doi.org/10.1103/PhysRevD.100.034027} {\bibfield  {journal} {\bibinfo
   {journal} {Phys. Rev. D}\ }\textbf {\bibinfo {volume} {100}},\ \bibinfo
  {pages} {034027} (\bibinfo {year} {2019})},\ \Eprint
  {https://arxiv.org/abs/1907.04475} {arXiv:1907.04475 [hep-ph]} \BibitemShut
  {NoStop}%
\bibitem [{\citenamefont {Miyahara}\ \emph {et~al.}(2017)\citenamefont
  {Miyahara}, \citenamefont {Hyodo}, \citenamefont {Oka}, \citenamefont
  {Nieves},\ and\ \citenamefont {Oset}}]{Miyahara:2016yyh}%
  \BibitemOpen
  \bibfield  {author} {\bibinfo {author} {\bibfnamefont {K.}~\bibnamefont
  {Miyahara}}, \bibinfo {author} {\bibfnamefont {T.}~\bibnamefont {Hyodo}},
  \bibinfo {author} {\bibfnamefont {M.}~\bibnamefont {Oka}}, \bibinfo {author}
  {\bibfnamefont {J.}~\bibnamefont {Nieves}},\ and\ \bibinfo {author}
  {\bibfnamefont {E.}~\bibnamefont {Oset}},\ }\bibfield  {title} {\bibinfo
  {title} {{Theoretical study of the \ensuremath{\Xi}(1620) and
  \ensuremath{\Xi}(1690) resonances in
  $\ensuremath{\Xi}_c\to\ensuremath{\pi}^+MB$ decays}},\ }\href
  {https://doi.org/10.1103/PhysRevC.95.035212} {\bibfield  {journal} {\bibinfo
  {journal} {Phys. Rev. C}\ }\textbf {\bibinfo {volume} {95}},\ \bibinfo
  {pages} {035212} (\bibinfo {year} {2017})},\ \Eprint
  {https://arxiv.org/abs/1609.00895} {arXiv:1609.00895 [nucl-th]} \BibitemShut
  {NoStop}%
\bibitem [{\citenamefont {Nieves}\ and\ \citenamefont
  {Ruiz~Arriola}(1999)}]{Nieves:1998hp}%
  \BibitemOpen
  \bibfield  {author} {\bibinfo {author} {\bibfnamefont {J.}~\bibnamefont
  {Nieves}}\ and\ \bibinfo {author} {\bibfnamefont {E.}~\bibnamefont
  {Ruiz~Arriola}},\ }\bibfield  {title} {\bibinfo {title} {{Bethe-Salpeter
  approach for meson meson scattering in chiral perturbation theory}},\ }\href
  {https://doi.org/10.1016/S0370-2693(99)00461-X} {\bibfield  {journal}
  {\bibinfo  {journal} {Phys. Lett. B}\ }\textbf {\bibinfo {volume} {455}},\
  \bibinfo {pages} {30} (\bibinfo {year} {1999})},\ \Eprint
  {https://arxiv.org/abs/nucl-th/9807035} {arXiv:nucl-th/9807035} \BibitemShut
  {NoStop}%
\bibitem [{\citenamefont {Oller}\ and\ \citenamefont
  {Oset}(1997)}]{Oller:1997ti}%
  \BibitemOpen
  \bibfield  {author} {\bibinfo {author} {\bibfnamefont {J.~A.}\ \bibnamefont
  {Oller}}\ and\ \bibinfo {author} {\bibfnamefont {E.}~\bibnamefont {Oset}},\
  }\bibfield  {title} {\bibinfo {title} {{Chiral symmetry amplitudes in the S
  wave isoscalar and isovector channels and the $\sigma$, $f_0$(980),
  $a_0$(980) scalar mesons}},\ }\href
  {https://doi.org/10.1016/S0375-9474(97)00160-7} {\bibfield  {journal}
  {\bibinfo  {journal} {Nucl. Phys. A}\ }\textbf {\bibinfo {volume} {620}},\
  \bibinfo {pages} {438} (\bibinfo {year} {1997})},\ \bibinfo {note} {[Erratum:
  Nucl.Phys.A 652, 407--409 (1999)]},\ \Eprint
  {https://arxiv.org/abs/hep-ph/9702314} {arXiv:hep-ph/9702314} \BibitemShut
  {NoStop}%
\bibitem [{\citenamefont {Kaiser}\ \emph {et~al.}(1995)\citenamefont {Kaiser},
  \citenamefont {Siegel},\ and\ \citenamefont {Weise}}]{Kaiser:1995cy}%
  \BibitemOpen
  \bibfield  {author} {\bibinfo {author} {\bibfnamefont {N.}~\bibnamefont
  {Kaiser}}, \bibinfo {author} {\bibfnamefont {P.~B.}\ \bibnamefont {Siegel}},\
  and\ \bibinfo {author} {\bibfnamefont {W.}~\bibnamefont {Weise}},\ }\bibfield
   {title} {\bibinfo {title} {{Chiral dynamics and the S$_{11}$ (1535) nucleon
  resonance}},\ }\href {https://doi.org/10.1016/0370-2693(95)01203-3}
  {\bibfield  {journal} {\bibinfo  {journal} {Phys. Lett. B}\ }\textbf
  {\bibinfo {volume} {362}},\ \bibinfo {pages} {23} (\bibinfo {year} {1995})},\
  \Eprint {https://arxiv.org/abs/nucl-th/9507036} {arXiv:nucl-th/9507036}
  \BibitemShut {NoStop}%
\bibitem [{\citenamefont {Nieves}\ and\ \citenamefont
  {Ruiz~Arriola}(2001)}]{Nieves:2001wt}%
  \BibitemOpen
  \bibfield  {author} {\bibinfo {author} {\bibfnamefont {J.}~\bibnamefont
  {Nieves}}\ and\ \bibinfo {author} {\bibfnamefont {E.}~\bibnamefont
  {Ruiz~Arriola}},\ }\bibfield  {title} {\bibinfo {title} {{The
  S$_{11}$-N(1535) and -N(1650) resonances in meson baryon unitarized coupled
  channel chiral perturbation theory}},\ }\href
  {https://doi.org/10.1103/PhysRevD.64.116008} {\bibfield  {journal} {\bibinfo
  {journal} {Phys. Rev. D}\ }\textbf {\bibinfo {volume} {64}},\ \bibinfo
  {pages} {116008} (\bibinfo {year} {2001})},\ \Eprint
  {https://arxiv.org/abs/hep-ph/0104307} {arXiv:hep-ph/0104307} \BibitemShut
  {NoStop}%
\bibitem [{\citenamefont {Inoue}\ \emph {et~al.}(2002)\citenamefont {Inoue},
  \citenamefont {Oset},\ and\ \citenamefont {Vicente~Vacas}}]{Inoue:2001ip}%
  \BibitemOpen
  \bibfield  {author} {\bibinfo {author} {\bibfnamefont {T.}~\bibnamefont
  {Inoue}}, \bibinfo {author} {\bibfnamefont {E.}~\bibnamefont {Oset}},\ and\
  \bibinfo {author} {\bibfnamefont {M.~J.}\ \bibnamefont {Vicente~Vacas}},\
  }\bibfield  {title} {\bibinfo {title} {{Chiral unitary approach to S wave
  meson baryon scattering in the strangeness S = O sector}},\ }\href
  {https://doi.org/10.1103/PhysRevC.65.035204} {\bibfield  {journal} {\bibinfo
  {journal} {Phys. Rev. C}\ }\textbf {\bibinfo {volume} {65}},\ \bibinfo
  {pages} {035204} (\bibinfo {year} {2002})},\ \Eprint
  {https://arxiv.org/abs/hep-ph/0110333} {arXiv:hep-ph/0110333} \BibitemShut
  {NoStop}%
\bibitem [{\citenamefont {Bruns}\ \emph {et~al.}(2011)\citenamefont {Bruns},
  \citenamefont {Mai},\ and\ \citenamefont {Meissner}}]{Bruns:2010sv}%
  \BibitemOpen
  \bibfield  {author} {\bibinfo {author} {\bibfnamefont {P.~C.}\ \bibnamefont
  {Bruns}}, \bibinfo {author} {\bibfnamefont {M.}~\bibnamefont {Mai}},\ and\
  \bibinfo {author} {\bibfnamefont {U.~G.}\ \bibnamefont {Meissner}},\
  }\bibfield  {title} {\bibinfo {title} {{Chiral dynamics of the S$_{11}$(1535)
  and S$_{11}$(1650) resonances revisited}},\ }\href
  {https://doi.org/10.1016/j.physletb.2011.02.008} {\bibfield  {journal}
  {\bibinfo  {journal} {Phys. Lett. B}\ }\textbf {\bibinfo {volume} {697}},\
  \bibinfo {pages} {254} (\bibinfo {year} {2011})},\ \Eprint
  {https://arxiv.org/abs/1012.2233} {arXiv:1012.2233 [nucl-th]} \BibitemShut
  {NoStop}%
\bibitem [{\citenamefont {Khemchandani}\ \emph {et~al.}(2013)\citenamefont
  {Khemchandani}, \citenamefont {Martinez~Torres}, \citenamefont {Nagahiro},\
  and\ \citenamefont {Hosaka}}]{Khemchandani:2013nma}%
  \BibitemOpen
  \bibfield  {author} {\bibinfo {author} {\bibfnamefont {K.~P.}\ \bibnamefont
  {Khemchandani}}, \bibinfo {author} {\bibfnamefont {A.}~\bibnamefont
  {Martinez~Torres}}, \bibinfo {author} {\bibfnamefont {H.}~\bibnamefont
  {Nagahiro}},\ and\ \bibinfo {author} {\bibfnamefont {A.}~\bibnamefont
  {Hosaka}},\ }\bibfield  {title} {\bibinfo {title} {{Role of vector and
  pseudoscalar mesons in understanding $1/2^- N^*$ and \ensuremath{\Delta}
  resonances}},\ }\href {https://doi.org/10.1103/PhysRevD.88.114016} {\bibfield
   {journal} {\bibinfo  {journal} {Phys. Rev. D}\ }\textbf {\bibinfo {volume}
  {88}},\ \bibinfo {pages} {114016} (\bibinfo {year} {2013})},\ \Eprint
  {https://arxiv.org/abs/1307.8420} {arXiv:1307.8420 [nucl-th]} \BibitemShut
  {NoStop}%
\bibitem [{\citenamefont {Garzon}\ and\ \citenamefont
  {Oset}(2015)}]{Garzon:2014ida}%
  \BibitemOpen
  \bibfield  {author} {\bibinfo {author} {\bibfnamefont {E.~J.}\ \bibnamefont
  {Garzon}}\ and\ \bibinfo {author} {\bibfnamefont {E.}~\bibnamefont {Oset}},\
  }\bibfield  {title} {\bibinfo {title} {{Mixing of pseudoscalar-baryon and
  vector-baryon in the $J^P=1/2^-$ sector and the $N^*$(1535) and $N^*$(1650)
  resonances}},\ }\href {https://doi.org/10.1103/PhysRevC.91.025201} {\bibfield
   {journal} {\bibinfo  {journal} {Phys. Rev. C}\ }\textbf {\bibinfo {volume}
  {91}},\ \bibinfo {pages} {025201} (\bibinfo {year} {2015})},\ \Eprint
  {https://arxiv.org/abs/1411.3547} {arXiv:1411.3547 [hep-ph]} \BibitemShut
  {NoStop}%
\bibitem [{\citenamefont {Xie}\ \emph {et~al.}(2015)\citenamefont {Xie},
  \citenamefont {Dai},\ and\ \citenamefont {Oset}}]{Xie:2014tma}%
  \BibitemOpen
  \bibfield  {author} {\bibinfo {author} {\bibfnamefont {J.-J.}\ \bibnamefont
  {Xie}}, \bibinfo {author} {\bibfnamefont {L.-R.}\ \bibnamefont {Dai}},\ and\
  \bibinfo {author} {\bibfnamefont {E.}~\bibnamefont {Oset}},\ }\bibfield
  {title} {\bibinfo {title} {{The low lying scalar resonances in the $D^0$
  decays into $K^0_S$ and $f_0(500)$, $f_0(980)$, $a_0(980)$}},\ }\href
  {https://doi.org/10.1016/j.physletb.2015.02.006} {\bibfield  {journal}
  {\bibinfo  {journal} {Phys. Lett. B}\ }\textbf {\bibinfo {volume} {742}},\
  \bibinfo {pages} {363} (\bibinfo {year} {2015})},\ \Eprint
  {https://arxiv.org/abs/1409.0401} {arXiv:1409.0401 [hep-ph]} \BibitemShut
  {NoStop}%
\bibitem [{\citenamefont {Ling}\ \emph {et~al.}(2021)\citenamefont {Ling},
  \citenamefont {Liu}, \citenamefont {Lu}, \citenamefont {Geng},\ and\
  \citenamefont {Xie}}]{Ling:2021qzl}%
  \BibitemOpen
  \bibfield  {author} {\bibinfo {author} {\bibfnamefont {X.-Z.}\ \bibnamefont
  {Ling}}, \bibinfo {author} {\bibfnamefont {M.-Z.}\ \bibnamefont {Liu}},
  \bibinfo {author} {\bibfnamefont {J.-X.}\ \bibnamefont {Lu}}, \bibinfo
  {author} {\bibfnamefont {L.-S.}\ \bibnamefont {Geng}},\ and\ \bibinfo
  {author} {\bibfnamefont {J.-J.}\ \bibnamefont {Xie}},\ }\bibfield  {title}
  {\bibinfo {title} {{Can the nature of $a_0(980)$ be tested in the $D_S^{+}\to
  \pi^{+}\pi^0 \eta$ decay?}},\ }\href
  {https://doi.org/10.1103/PhysRevD.103.116016} {\bibfield  {journal} {\bibinfo
   {journal} {Phys. Rev. D}\ }\textbf {\bibinfo {volume} {103}},\ \bibinfo
  {pages} {116016} (\bibinfo {year} {2021})},\ \Eprint
  {https://arxiv.org/abs/2102.05349} {arXiv:2102.05349 [hep-ph]} \BibitemShut
  {NoStop}%
\bibitem [{\citenamefont {Wang}\ \emph {et~al.}(2021)\citenamefont {Wang},
  \citenamefont {Duan}, \citenamefont {Wang}, \citenamefont {Li}, \citenamefont
  {Liu},\ and\ \citenamefont {Wang}}]{Wang:2021naf}%
  \BibitemOpen
  \bibfield  {author} {\bibinfo {author} {\bibfnamefont {J.-Y.}\ \bibnamefont
  {Wang}}, \bibinfo {author} {\bibfnamefont {M.-Y.}\ \bibnamefont {Duan}},
  \bibinfo {author} {\bibfnamefont {G.-Y.}\ \bibnamefont {Wang}}, \bibinfo
  {author} {\bibfnamefont {D.-M.}\ \bibnamefont {Li}}, \bibinfo {author}
  {\bibfnamefont {L.-J.}\ \bibnamefont {Liu}},\ and\ \bibinfo {author}
  {\bibfnamefont {E.}~\bibnamefont {Wang}},\ }\bibfield  {title} {\bibinfo
  {title} {{The $a_0(980)$ and $f_0(980)$ in the process $D_S^+\to
  K^+K^-\ensuremath{\pi}^+$}},\ }\href
  {https://doi.org/10.1016/j.physletb.2021.136617} {\bibfield  {journal}
  {\bibinfo  {journal} {Phys. Lett. B}\ }\textbf {\bibinfo {volume} {821}},\
  \bibinfo {pages} {136617} (\bibinfo {year} {2021})},\ \Eprint
  {https://arxiv.org/abs/2105.04907} {arXiv:2105.04907 [hep-ph]} \BibitemShut
  {NoStop}%
\bibitem [{\citenamefont {Molina}\ \emph {et~al.}(2020)\citenamefont {Molina},
  \citenamefont {Xie}, \citenamefont {Liang}, \citenamefont {Geng},\ and\
  \citenamefont {Oset}}]{Molina:2019udw}%
  \BibitemOpen
  \bibfield  {author} {\bibinfo {author} {\bibfnamefont {R.}~\bibnamefont
  {Molina}}, \bibinfo {author} {\bibfnamefont {J.-J.}\ \bibnamefont {Xie}},
  \bibinfo {author} {\bibfnamefont {W.-H.}\ \bibnamefont {Liang}}, \bibinfo
  {author} {\bibfnamefont {L.-S.}\ \bibnamefont {Geng}},\ and\ \bibinfo
  {author} {\bibfnamefont {E.}~\bibnamefont {Oset}},\ }\bibfield  {title}
  {\bibinfo {title} {{Theoretical interpretation of the $D^+_S \to \pi^+ \pi^0
  \eta$ decay and the nature of $a_0(980)$}},\ }\href
  {https://doi.org/10.1016/j.physletb.2020.135279} {\bibfield  {journal}
  {\bibinfo  {journal} {Phys. Lett. B}\ }\textbf {\bibinfo {volume} {803}},\
  \bibinfo {pages} {135279} (\bibinfo {year} {2020})},\ \Eprint
  {https://arxiv.org/abs/1908.11557} {arXiv:1908.11557 [hep-ph]} \BibitemShut
  {NoStop}%
\bibitem [{\citenamefont {Duan}\ \emph {et~al.}(2020)\citenamefont {Duan},
  \citenamefont {Wang}, \citenamefont {Wang}, \citenamefont {Wang},\ and\
  \citenamefont {Li}}]{Duan:2020vye}%
  \BibitemOpen
  \bibfield  {author} {\bibinfo {author} {\bibfnamefont {M.-Y.}\ \bibnamefont
  {Duan}}, \bibinfo {author} {\bibfnamefont {J.-Y.}\ \bibnamefont {Wang}},
  \bibinfo {author} {\bibfnamefont {G.-Y.}\ \bibnamefont {Wang}}, \bibinfo
  {author} {\bibfnamefont {E.}~\bibnamefont {Wang}},\ and\ \bibinfo {author}
  {\bibfnamefont {D.-M.}\ \bibnamefont {Li}},\ }\bibfield  {title} {\bibinfo
  {title} {{Role of scalar $a_0(980)$ in the single Cabibbo suppressed process
  $D^+ \rightarrow \pi ^{+} \pi ^{0} \eta $}},\ }\href
  {https://doi.org/10.1140/epjc/s10052-020-08630-3} {\bibfield  {journal}
  {\bibinfo  {journal} {Eur. Phys. J. C}\ }\textbf {\bibinfo {volume} {80}},\
  \bibinfo {pages} {1041} (\bibinfo {year} {2020})},\ \Eprint
  {https://arxiv.org/abs/2008.10139} {arXiv:2008.10139 [hep-ph]} \BibitemShut
  {NoStop}%
\bibitem [{\citenamefont {Zhu}\ \emph {et~al.}(2022)\citenamefont {Zhu},
  \citenamefont {Wang}, \citenamefont {Li}, \citenamefont {Wang}, \citenamefont
  {Geng},\ and\ \citenamefont {Xie}}]{Zhu:2022duu}%
  \BibitemOpen
  \bibfield  {author} {\bibinfo {author} {\bibfnamefont {X.}~\bibnamefont
  {Zhu}}, \bibinfo {author} {\bibfnamefont {H.-N.}\ \bibnamefont {Wang}},
  \bibinfo {author} {\bibfnamefont {D.-M.}\ \bibnamefont {Li}}, \bibinfo
  {author} {\bibfnamefont {E.}~\bibnamefont {Wang}}, \bibinfo {author}
  {\bibfnamefont {L.-S.}\ \bibnamefont {Geng}},\ and\ \bibinfo {author}
  {\bibfnamefont {J.-J.}\ \bibnamefont {Xie}},\ }\bibfield  {title} {\bibinfo
  {title} {{Further understanding the nature of $a_0(1710)$ in the $D^+_S \to
  \pi^0 K^+ K^0_S$ decay}},\ }\href@noop {} {\  (\bibinfo {year} {2022})},\
  \Eprint {https://arxiv.org/abs/2210.12992} {arXiv:2210.12992 [hep-ph]}
  \BibitemShut {NoStop}%
\bibitem [{\citenamefont {Xie}\ and\ \citenamefont {Geng}(2016)}]{Xie:2016evi}%
  \BibitemOpen
  \bibfield  {author} {\bibinfo {author} {\bibfnamefont {J.-J.}\ \bibnamefont
  {Xie}}\ and\ \bibinfo {author} {\bibfnamefont {L.-S.}\ \bibnamefont {Geng}},\
  }\bibfield  {title} {\bibinfo {title} {{The $a_0(980)$ and $\Lambda(1670)$ in
  the $\Lambda^+_c \to \pi^+ \eta \Lambda$ decay}},\ }\href
  {https://doi.org/10.1140/epjc/s10052-016-4342-z} {\bibfield  {journal}
  {\bibinfo  {journal} {Eur. Phys. J. C}\ }\textbf {\bibinfo {volume} {76}},\
  \bibinfo {pages} {496} (\bibinfo {year} {2016})},\ \Eprint
  {https://arxiv.org/abs/1604.02756} {arXiv:1604.02756 [nucl-th]} \BibitemShut
  {NoStop}%
\bibitem [{\citenamefont {Xie}\ and\ \citenamefont
  {Geng}(2017{\natexlab{a}})}]{Xie:2017erh}%
  \BibitemOpen
  \bibfield  {author} {\bibinfo {author} {\bibfnamefont {J.-J.}\ \bibnamefont
  {Xie}}\ and\ \bibinfo {author} {\bibfnamefont {L.-S.}\ \bibnamefont {Geng}},\
  }\bibfield  {title} {\bibinfo {title} {{Role of the $N^*(1535)$ in the
  $\Lambda^+_c \to \bar{K}^0 \eta p$ decay}},\ }\href
  {https://doi.org/10.1103/PhysRevD.96.054009} {\bibfield  {journal} {\bibinfo
  {journal} {Phys. Rev. D}\ }\textbf {\bibinfo {volume} {96}},\ \bibinfo
  {pages} {054009} (\bibinfo {year} {2017}{\natexlab{a}})},\ \Eprint
  {https://arxiv.org/abs/1704.05714} {arXiv:1704.05714 [hep-ph]} \BibitemShut
  {NoStop}%
\bibitem [{\citenamefont {Xie}\ and\ \citenamefont
  {Geng}(2017{\natexlab{b}})}]{Xie:2017xwx}%
  \BibitemOpen
  \bibfield  {author} {\bibinfo {author} {\bibfnamefont {J.-J.}\ \bibnamefont
  {Xie}}\ and\ \bibinfo {author} {\bibfnamefont {L.-S.}\ \bibnamefont {Geng}},\
  }\bibfield  {title} {\bibinfo {title} {{$\Sigma^*_{1/2^-}(1380)$ in the
  $\Lambda^+_c \to \eta \pi^+ \Lambda$ decay}},\ }\href
  {https://doi.org/10.1103/PhysRevD.95.074024} {\bibfield  {journal} {\bibinfo
  {journal} {Phys. Rev. D}\ }\textbf {\bibinfo {volume} {95}},\ \bibinfo
  {pages} {074024} (\bibinfo {year} {2017}{\natexlab{b}})},\ \Eprint
  {https://arxiv.org/abs/1703.09502} {arXiv:1703.09502 [hep-ph]} \BibitemShut
  {NoStop}%
\bibitem [{\citenamefont {Wang}\ \emph {et~al.}(2022)\citenamefont {Wang},
  \citenamefont {Wei}, \citenamefont {Yang}, \citenamefont {Wang},
  \citenamefont {Geng},\ and\ \citenamefont {Xie}}]{Wang:2022nac}%
  \BibitemOpen
  \bibfield  {author} {\bibinfo {author} {\bibfnamefont {G.-Y.}\ \bibnamefont
  {Wang}}, \bibinfo {author} {\bibfnamefont {N.-C.}\ \bibnamefont {Wei}},
  \bibinfo {author} {\bibfnamefont {H.-M.}\ \bibnamefont {Yang}}, \bibinfo
  {author} {\bibfnamefont {E.}~\bibnamefont {Wang}}, \bibinfo {author}
  {\bibfnamefont {L.-S.}\ \bibnamefont {Geng}},\ and\ \bibinfo {author}
  {\bibfnamefont {J.-J.}\ \bibnamefont {Xie}},\ }\bibfield  {title} {\bibinfo
  {title} {{Roles of $a_0(980)$, \ensuremath{\Lambda}(1670), and
  \ensuremath{\Sigma}(1385) in the
  $\ensuremath{\Lambda}_c^+\to\ensuremath{\eta}\ensuremath{\Lambda}\ensuremath{\pi}^+$
  decay}},\ }\href {https://doi.org/10.1103/PhysRevD.106.056001} {\bibfield
  {journal} {\bibinfo  {journal} {Phys. Rev. D}\ }\textbf {\bibinfo {volume}
  {106}},\ \bibinfo {pages} {056001} (\bibinfo {year} {2022})},\ \Eprint
  {https://arxiv.org/abs/2206.01425} {arXiv:2206.01425 [hep-ph]} \BibitemShut
  {NoStop}%
\bibitem [{\citenamefont {Pavao}\ \emph {et~al.}(2018)\citenamefont {Pavao},
  \citenamefont {Sakai},\ and\ \citenamefont {Oset}}]{Pavao:2018wdf}%
  \BibitemOpen
  \bibfield  {author} {\bibinfo {author} {\bibfnamefont {R.}~\bibnamefont
  {Pavao}}, \bibinfo {author} {\bibfnamefont {S.}~\bibnamefont {Sakai}},\ and\
  \bibinfo {author} {\bibfnamefont {E.}~\bibnamefont {Oset}},\ }\bibfield
  {title} {\bibinfo {title} {{Production of $N^*(1535)$ and $N^*(1650)$ in
  $\Lambda_c\rightarrow\bar{K}^0\eta p$ $(\pi N)$ decay}},\ }\href
  {https://doi.org/10.1103/PhysRevC.98.015201} {\bibfield  {journal} {\bibinfo
  {journal} {Phys. Rev. C}\ }\textbf {\bibinfo {volume} {98}},\ \bibinfo
  {pages} {015201} (\bibinfo {year} {2018})},\ \Eprint
  {https://arxiv.org/abs/1802.07882} {arXiv:1802.07882 [nucl-th]} \BibitemShut
  {NoStop}%
\bibitem [{\citenamefont {Chau}\ and\ \citenamefont
  {Cheng}(1987)}]{Chau:1987tk}%
  \BibitemOpen
  \bibfield  {author} {\bibinfo {author} {\bibfnamefont {L.-L.}\ \bibnamefont
  {Chau}}\ and\ \bibinfo {author} {\bibfnamefont {H.-Y.}\ \bibnamefont
  {Cheng}},\ }\bibfield  {title} {\bibinfo {title} {{Analysis of Exclusive
  Two-Body Decays of Charm Mesons Using the Quark Diagram Scheme}},\ }\href
  {https://doi.org/10.1103/PhysRevD.39.2788} {\bibfield  {journal} {\bibinfo
  {journal} {Phys. Rev. D}\ }\textbf {\bibinfo {volume} {36}},\ \bibinfo
  {pages} {137} (\bibinfo {year} {1987})},\ \bibinfo {note} {[Addendum:
  Phys.Rev.D 39, 2788--2791 (1989)]}\BibitemShut {NoStop}%
\bibitem [{\citenamefont {Chau}\ and\ \citenamefont
  {Cheng}(1990)}]{Chau:1990ij}%
  \BibitemOpen
  \bibfield  {author} {\bibinfo {author} {\bibfnamefont {L.~L.}\ \bibnamefont
  {Chau}}\ and\ \bibinfo {author} {\bibfnamefont {H.-Y.}\ \bibnamefont
  {Cheng}},\ }\bibfield  {title} {\bibinfo {title} {{Nonresonant three-body
  decays of charmed mesons}},\ }\href
  {https://doi.org/10.1103/PhysRevD.41.1510} {\bibfield  {journal} {\bibinfo
  {journal} {Phys. Rev. D}\ }\textbf {\bibinfo {volume} {41}},\ \bibinfo
  {pages} {1510} (\bibinfo {year} {1990})}\BibitemShut {NoStop}%
\bibitem [{\citenamefont {Chau}\ \emph {et~al.}(1996)\citenamefont {Chau},
  \citenamefont {Cheng},\ and\ \citenamefont {Tseng}}]{Chau:1995gk}%
  \BibitemOpen
  \bibfield  {author} {\bibinfo {author} {\bibfnamefont {L.-L.}\ \bibnamefont
  {Chau}}, \bibinfo {author} {\bibfnamefont {H.-Y.}\ \bibnamefont {Cheng}},\
  and\ \bibinfo {author} {\bibfnamefont {B.}~\bibnamefont {Tseng}},\ }\bibfield
   {title} {\bibinfo {title} {{Analysis of two-body decays of charmed baryons
  using the quark diagram scheme}},\ }\href
  {https://doi.org/10.1103/PhysRevD.54.2132} {\bibfield  {journal} {\bibinfo
  {journal} {Phys. Rev. D}\ }\textbf {\bibinfo {volume} {54}},\ \bibinfo
  {pages} {2132} (\bibinfo {year} {1996})},\ \Eprint
  {https://arxiv.org/abs/hep-ph/9508382} {arXiv:hep-ph/9508382} \BibitemShut
  {NoStop}%
\bibitem [{\citenamefont {Qin}\ \emph {et~al.}(2022)\citenamefont {Qin},
  \citenamefont {Qiu},\ and\ \citenamefont {Yu}}]{Qin:2022nof}%
  \BibitemOpen
  \bibfield  {author} {\bibinfo {author} {\bibfnamefont {Q.}~\bibnamefont
  {Qin}}, \bibinfo {author} {\bibfnamefont {J.-L.}\ \bibnamefont {Qiu}},\ and\
  \bibinfo {author} {\bibfnamefont {F.-S.}\ \bibnamefont {Yu}},\ }\bibfield
  {title} {\bibinfo {title} {{Diagrammatic analysis of hidden- and open-charm
  tetraquark production in $B$ decays}},\ }\href@noop {} {\  (\bibinfo {year}
  {2022})},\ \Eprint {https://arxiv.org/abs/2212.03590} {arXiv:2212.03590
  [hep-ph]} \BibitemShut {NoStop}%
\bibitem [{\citenamefont {Oller}\ \emph {et~al.}(1998)\citenamefont {Oller},
  \citenamefont {Oset},\ and\ \citenamefont {Pelaez}}]{Oller:1997ng}%
  \BibitemOpen
  \bibfield  {author} {\bibinfo {author} {\bibfnamefont {J.~A.}\ \bibnamefont
  {Oller}}, \bibinfo {author} {\bibfnamefont {E.}~\bibnamefont {Oset}},\ and\
  \bibinfo {author} {\bibfnamefont {J.~R.}\ \bibnamefont {Pelaez}},\ }\bibfield
   {title} {\bibinfo {title} {{Nonperturbative approach to effective chiral
  Lagrangians and meson interactions}},\ }\href
  {https://doi.org/10.1103/PhysRevLett.80.3452} {\bibfield  {journal} {\bibinfo
   {journal} {Phys. Rev. Lett.}\ }\textbf {\bibinfo {volume} {80}},\ \bibinfo
  {pages} {3452} (\bibinfo {year} {1998})},\ \Eprint
  {https://arxiv.org/abs/hep-ph/9803242} {arXiv:hep-ph/9803242} \BibitemShut
  {NoStop}%
\bibitem [{\citenamefont {Oller}\ \emph {et~al.}(1999)\citenamefont {Oller},
  \citenamefont {Oset},\ and\ \citenamefont {Pelaez}}]{Oller:1998hw}%
  \BibitemOpen
  \bibfield  {author} {\bibinfo {author} {\bibfnamefont {J.~A.}\ \bibnamefont
  {Oller}}, \bibinfo {author} {\bibfnamefont {E.}~\bibnamefont {Oset}},\ and\
  \bibinfo {author} {\bibfnamefont {J.~R.}\ \bibnamefont {Pelaez}},\ }\bibfield
   {title} {\bibinfo {title} {{Meson meson interaction in a nonperturbative
  chiral approach}},\ }\href {https://doi.org/10.1103/PhysRevD.59.074001}
  {\bibfield  {journal} {\bibinfo  {journal} {Phys. Rev. D}\ }\textbf {\bibinfo
  {volume} {59}},\ \bibinfo {pages} {074001} (\bibinfo {year} {1999})},\
  \bibinfo {note} {[Erratum: Phys.Rev.D 60, 099906 (1999), Erratum: Phys.Rev.D
  75, 099903 (2007)]},\ \Eprint {https://arxiv.org/abs/hep-ph/9804209}
  {arXiv:hep-ph/9804209} \BibitemShut {NoStop}%
\bibitem [{\citenamefont {Garzon}\ and\ \citenamefont
  {Oset}(2016)}]{Garzon:2016evl}%
  \BibitemOpen
  \bibfield  {author} {\bibinfo {author} {\bibfnamefont {E.~J.}\ \bibnamefont
  {Garzon}}\ and\ \bibinfo {author} {\bibfnamefont {E.}~\bibnamefont {Oset}},\
  }\bibfield  {title} {\bibinfo {title} {{Mixing of Pseudoscalar-Baryon and
  Vector-Baryon in the $J^P = 1/2^-$ Sector and the $N^*(1535)$ and $N^*(1650)$
  resonances }},\ }\href {https://doi.org/10.7566/JPSCP.10.022004} {\bibfield
  {journal} {\bibinfo  {journal} {JPS Conf. Proc.}\ }\textbf {\bibinfo {volume}
  {10}},\ \bibinfo {pages} {022004} (\bibinfo {year} {2016})}\BibitemShut
  {NoStop}%
\bibitem [{\citenamefont {Ramos}\ \emph {et~al.}(2002)\citenamefont {Ramos},
  \citenamefont {Oset},\ and\ \citenamefont {Bennhold}}]{Ramos:2002xh}%
  \BibitemOpen
  \bibfield  {author} {\bibinfo {author} {\bibfnamefont {A.}~\bibnamefont
  {Ramos}}, \bibinfo {author} {\bibfnamefont {E.}~\bibnamefont {Oset}},\ and\
  \bibinfo {author} {\bibfnamefont {C.}~\bibnamefont {Bennhold}},\ }\bibfield
  {title} {\bibinfo {title} {{On the spin, parity and nature of the $\Xi(1620)$
  resonance}},\ }\href {https://doi.org/10.1103/PhysRevLett.89.252001}
  {\bibfield  {journal} {\bibinfo  {journal} {Phys. Rev. Lett.}\ }\textbf
  {\bibinfo {volume} {89}},\ \bibinfo {pages} {252001} (\bibinfo {year}
  {2002})},\ \Eprint {https://arxiv.org/abs/nucl-th/0204044}
  {arXiv:nucl-th/0204044} \BibitemShut {NoStop}%
\bibitem [{\citenamefont {Avery}\ \emph {et~al.}(1993)\citenamefont {Avery}
  \emph {et~al.}}]{CLEO:1993fhs}%
  \BibitemOpen
  \bibfield  {author} {\bibinfo {author} {\bibfnamefont {P.}~\bibnamefont
  {Avery}} \emph {et~al.} (\bibinfo {collaboration} {CLEO}),\ }\bibfield
  {title} {\bibinfo {title} {{Study of the decays $\Lambda_c^+ \to \Xi^0 K^+$,
  $\Lambda_c^+ \to \Sigma^+ K^+ K^-$ and $\Lambda_c^+ \to \Xi^- K^+ \pi^+$}},\
  }\href {https://doi.org/10.1103/PhysRevLett.71.2391} {\bibfield  {journal}
  {\bibinfo  {journal} {Phys. Rev. Lett.}\ }\textbf {\bibinfo {volume} {71}},\
  \bibinfo {pages} {2391} (\bibinfo {year} {1993})}\BibitemShut {NoStop}%
\bibitem [{\citenamefont {Ablikim}\ \emph {et~al.}(2018)\citenamefont {Ablikim}
  \emph {et~al.}}]{BESIII:2018cvs}%
  \BibitemOpen
  \bibfield  {author} {\bibinfo {author} {\bibfnamefont {M.}~\bibnamefont
  {Ablikim}} \emph {et~al.} (\bibinfo {collaboration} {BESIII}),\ }\bibfield
  {title} {\bibinfo {title} {{Measurements of absolute branching fractions for
  $\Lambda^+_c\to\Xi^0K^+$ and $\Xi(1530)^0K^+$}},\ }\href
  {https://doi.org/10.1016/j.physletb.2018.06.046} {\bibfield  {journal}
  {\bibinfo  {journal} {Phys. Lett. B}\ }\textbf {\bibinfo {volume} {783}},\
  \bibinfo {pages} {200} (\bibinfo {year} {2018})},\ \Eprint
  {https://arxiv.org/abs/1803.04299} {arXiv:1803.04299 [hep-ex]} \BibitemShut
  {NoStop}%
\bibitem [{\citenamefont {Gronau}\ \emph {et~al.}(2018)\citenamefont {Gronau},
  \citenamefont {Rosner},\ and\ \citenamefont {Wohl}}]{Gronau:2018vei}%
  \BibitemOpen
  \bibfield  {author} {\bibinfo {author} {\bibfnamefont {M.}~\bibnamefont
  {Gronau}}, \bibinfo {author} {\bibfnamefont {J.~L.}\ \bibnamefont {Rosner}},\
  and\ \bibinfo {author} {\bibfnamefont {C.~G.}\ \bibnamefont {Wohl}},\
  }\bibfield  {title} {\bibinfo {title} {{Overview of $\Lambda_c$ decays}},\
  }\href {https://doi.org/10.1103/PhysRevD.97.116015} {\bibfield  {journal}
  {\bibinfo  {journal} {Phys. Rev. D}\ }\textbf {\bibinfo {volume} {97}},\
  \bibinfo {pages} {116015} (\bibinfo {year} {2018})},\ \bibinfo {note}
  {[Addendum: Phys.Rev.D 98, 073003 (2018)]},\ \Eprint
  {https://arxiv.org/abs/1808.03720} {arXiv:1808.03720 [hep-ph]} \BibitemShut
  {NoStop}%
\bibitem [{\citenamefont {Ablikim}\ \emph {et~al.}(2023)\citenamefont {Ablikim}
  \emph {et~al.}}]{BESIII:2023rky}%
  \BibitemOpen
  \bibfield  {author} {\bibinfo {author} {\bibfnamefont {M.}~\bibnamefont
  {Ablikim}} \emph {et~al.} (\bibinfo {collaboration} {BESIII}),\ }\bibfield
  {title} {\bibinfo {title} {{Measurement of branching fractions of
  $\Lambda_{c}^{+}$ decays to $\Sigma^{+} K^{+} K^{-}$, $\Sigma^{+}\phi$ and
  $\Sigma^{+} K^{+} \pi^{-}(\pi^{0})$}},\ }\href@noop {} {\  (\bibinfo {year}
  {2023})},\ \Eprint {https://arxiv.org/abs/2304.09405} {arXiv:2304.09405
  [hep-ex]} \BibitemShut {NoStop}%
\end{thebibliography}%

\end{document}